\documentclass[fleqn,usenatbib]{mnras}

\usepackage{newtxtext,newtxmath}

\usepackage[T1]{fontenc}

\DeclareRobustCommand{\VAN}[3]{#2}
\let\VANthebibliography\thebibliography
\def\thebibliography{\DeclareRobustCommand{\VAN}[3]{##3}\VANthebibliography}

\usepackage{graphicx}	
\usepackage{amsmath}

\usepackage[export]{adjustbox}

\usepackage{xcolor}

\newcommand*{\mysub}[2]{\ensuremath{#1_{\mathrm{#2}}}}
\newcommand*{\Omegam}{\mysub{\Omega}{m}}
\newcommand*{\Omegal}{\mysub{\Omega}{\Lambda}}
\newcommand*{\LCDM}{$\Lambda$CDM}
\newcommand*{\NH}{\mysub{N}{H}}
\newcommand*{\nelec}{\mysub{n}{e}}
\newcommand*{\nH}{\mysub{n}{H}}
\newcommand*{\dL}{\mysub{d}{L}}
\newcommand*{\dA}{\mysub{d}{A}}
\newcommand*{\dref}{\ensuremath{d_\mathrm{A}^\mathrm{ref}}}
\newcommand*{\rxs}{\ensuremath{\mathcal{R}}}
\newcommand*{\sint}{\mysub{\sigma}{int}}
\newcommand*{\smean}{\ensuremath{\overline{\ln\rxs}}}
\newcommand*{\bkt}{\mysub{b}{T}}
\newcommand*{\bF}{\mysub{b}{f}}
\newcommand*{\rwlx}{\mysub{R}{wl-x}}
\newcommand*{\bhse}{\mysub{b}{HSE}}
\newcommand*{\gtsim}{\ {\raise-.75ex\hbox{$\buildrel>\over\sim$}}\ }
\newcommand*{\ltsim}{\ {\raise-.75ex\hbox{$\buildrel<\over\sim$}}\ }
\newcommand*{\kmsMpc}{\ensuremath{\mathrm{km}\,\mathrm{s}^{-1}\,\mathrm{Mpc}^{-1}}}

\title[Measuring $H_0$ with dynamically relaxed clusters]{Measuring $H_0$ using X-ray and SZ effect observations of dynamically relaxed galaxy clusters}

\author[Wan et al.]{%
Jenny T. Wan,$^{1}$\thanks{E-mail: jwan@caltech.edu}
Adam B. Mantz,$^{2}$
Jack Sayers,$^{1}$
Steven W. Allen,$^{2,3,4}$\newauthor
R. Glenn Morris,$^{2,4}$
and Sunil R. Golwala$^{1}$
\\
$^{1}$Division of Physics, Mathematics and Astronomy, California Institute of Technology, 1200 E. California Blvd., Pasadena, CA 91125, USA\\
$^{2}$Kavli Institute for Particle Astrophysics and Cosmology, Stanford University, 452 Lomita Mall, Stanford, CA 94305, USA\\
$^{3}$Department of Physics, Stanford University, 382 Via Pueblo Mall, Stanford, CA 94305, USA\\
$^{4}$SLAC National Accelerator Laboratory, 2575 Sand Hill Road, Menlo Park, CA  94025, USA
}

\date{Accepted XXX. Received YYY; in original form ZZZ}

\pubyear{2020}

\begin{document}
\label{firstpage}
\pagerange{\pageref{firstpage}--\pageref{lastpage}}
\maketitle

\begin{abstract}
We use a sample of 14 massive, dynamically relaxed galaxy clusters to constrain the Hubble Constant, $H_0$, by combining X-ray and Sunyaev-Zel'dovich (SZ) effect signals measured with {\it Chandra}, {\it Planck} and Bolocam.
This is the first such analysis to marginalize over an empirical, data-driven prior on the overall accuracy of X-ray temperature measurements, while our restriction to the most relaxed, massive clusters also minimizes astrophysical systematics.
For a cosmological-constant model with $\Omegam = 0.3$ and $\Omegal = 0.7$, we find $H_0 = 67.3^{+21.3}_{-13.3}\,\kmsMpc$, limited by the temperature calibration uncertainty (compared to the statistically limited constraint of $H_0 = 72.3^{+7.6}_{-7.6}\,\kmsMpc$).
The intrinsic scatter in the X-ray/SZ pressure ratio is found to be $13 \pm 4$ per cent ($10 \pm 3$ per cent when two clusters with significant galactic dust emission are removed from the sample), 
consistent with being primarily due to triaxiality and projection.
We discuss the prospects for reducing the dominant systematic limitation to this analysis, with improved X-ray calibration and/or precise measurements of the relativistic SZ effect providing a plausible route to per cent level constraints on $H_0$.
\end{abstract}

\begin{keywords}
 cosmological parameters -- cosmology: observations -- distance scale -- galaxies: clusters: general -- X-rays: galaxies: clusters
\end{keywords}

\section{Introduction}

The universe is expanding. The present day rate of this expansion is known as the Hubble Constant, $H_0$, and its value has been debated for the past century \citep[e.g.,][]{Livio2013, You2017}.
Recent measurements based on the primary Cosmic Microwave Background (CMB) anisotropies observed by {\it Planck} indicate a value of $H_0 = 67.66 \pm 0.42$\,\kmsMpc{} \citep{Planck2018_VI}.
This precise constraint follows from the assumption of a cosmological-constant model and global spatial flatness (flat \LCDM{}), but holds with slightly larger uncertainties ($\sim1$\,\kmsMpc) for more flexible cosmological models when combined with independent constraints on the cosmic expansion history.
Combinations of Baryon Acoustic Oscillation (BAO) data with external priors tend to produce consistent values (e.g., \citealt{Aubourg2015, Addison2018}).
Since the BAO scale is determined by the sound horizon at recombination, BAO and CMB data have frequently been described as ``early Universe'' probes of $H_0$, in contrast to ``late Universe'' probes.
Most salient among the latter is the classic distance-ladder method, calibrated via the Cepheid period--luminosity relation, which yields a significantly higher value of $H_0 = 74.22 \pm 1.82$\,\kmsMpc \citep{Riess2019}.
Estimates from other ``late'' probes, including alternative calibration of the distance ladder using the tip of the red giant branch \citep{Freedman2020}, gravitational lensing time delays (e.g., \citealt{Birrer2020}), distance measurements based on megamasers \citep{Pesce2020}, gravitational wave sirens \citep{Abbott2017}, and combination of the gas mass fraction of massive clusters with the cosmic baryon fraction from the CMB (e.g., \citealt{Mantz1402.6212}), tend to fall between these extremes (and are frequently consistent with both).
Because an unambiguous early-late dichotomy in the inferred value of $H_0$ could be an indication of new physics \citep[e.g.,][]{Poulin2019}, the current state of the field motivates careful consideration of many independent techniques.

The combination of X-ray imaging spectroscopy and mm-wavelength observations of galaxy clusters provides one such independent avenue to probe the expansion of the Universe \citep[e.g.,][]{Silk1978}. The inverse-Compton scattering of CMB photons by thermal electrons in the hot intracluster medium (ICM) leads to a small ($\sim1$\,mK) distortion of the CMB spectrum along the line of sight through a cluster, known as the thermal Sunyaev-Zel'dovich (SZ) effect \citep{Sunyaev1972}.
The magnitude of the SZ effect is quantified by the Compton $y$ parameter, which is proportional to the line-of-sight integral of the electron pressure. The same hot ICM radiates X-rays through a combination of thermal bremsstrahlung and line emission, which can be used to independently probe the ICM pressure. 
The combination of X-ray and SZ measurements, along with a model of the cluster gas geometry, enables a direct determination of the distance to a cluster, independent of the extragalactic distance ladder \citep[e.g.,][]{Silk1978,Birkinshaw1979, Birkinshaw1991, Reese2000}. 

Interest in this technique grew when the requisite SZ effect and X-ray data became available for modest samples of clusters in the late 1990's and early 2000's \citep{Mason2001, Reese2002, Schmidt2004, Jones2005, Bonamente2006}.
However, several important systematic effects have limited the precision of $H_0$ estimates from these data.
Foremost among these is the need for an accurate absolute calibration of ICM temperatures measured from X-ray data, a challenging prospect \citep{Nevalainen2010, Schellenberger2015}.
\citet{Kozmanyan2019} recently performed an analysis that did not account for this critical source of uncertainty, reporting a tight constraint on $H_0$ at the $\pm3$\,\kmsMpc{} level,  demonstrating the statistical power currently available if systematics can be controlled.
Another challenge, particularly for analyses where the clusters are unresolved, is the presence of complex thermodynamic structure in merging clusters,
as well as the possibility of biases in X-ray measurements of gas density at large radii due to clumping \citep[e.g.,][]{Simionescu2011, Urban2014, Eckert2015}.

In this work, we provide an updated constraint on $H_0$ from X-ray and SZ cluster data, for the first time marginalizing over an empirical, data-driven prior on the overall accuracy of X-ray temperature measurements.
We minimize the impact of other systematics by restricting the analysis to the most dynamically relaxed, and hence geometrically simple, clusters known, for which high-quality, resolved X-ray and SZ data are available.
This choice allows us to make measurements at intermediate cluster radii, with minimal sensitivity to feedback from Active Galactic Nuclei (AGN) near the clusters' centers or ongoing accretion and gas clumping in the clusters' outskirts.

The structure of this paper is as follows. In Section~\ref{sec:data}, we present the X-ray and SZ data used for this analysis. We describe our methods for measuring the ICM pressure using these data, and extracting cosmological constraints, in Section~\ref{sec:methods}. Section~\ref{sec:results} presents our results and compares them to others in the literature, and Section~\ref{sec:future} discusses the prospects for improvements in the future. We summarize our conclusions in Section~\ref{sec:conclusion}.

\section{Data/Observations} \label{sec:data}

\subsection{Cluster sample}
\label{sec:sample}

We initially consider the sample of 40 hot ($kT\gtsim5$\,keV), dynamically relaxed clusters identified by \citet{Mantz1402.6212, Mantz1502.06020} using their Symmetry-Peakiness-Alignment (SPA) criterion.
While dynamical relaxation is not per se a requirement for this analysis, the morphological regularity and absence of thermodynamic substructure that are features of relaxed clusters minimize systematic scatter associated with the estimation of 3-dimensional thermodynamic profiles from projected X-ray and SZ effect data.
Note that the SPA selection algorithm does not advantage clusters that appear circular as opposed to ellipsoidal in projection, an important consideration in the present context, as even the most relaxed clusters are expected to be triaxial in general \citep[e.g.,][]{Jing2002}.
Hence, those presenting the smallest ellipticities in the plane of the sky may well have a significantly different extent along the line of sight (see Section~\ref{sec:asphericity}).

The SPA-selected sample uniformly has adequate {\it Chandra} X-ray data for the determination of gas density and temperature profiles \citep{Mantz1509.01322}.
A further selection is made based on the availability of SZ effect data. In order to perform the deprojection analysis described in Section~\ref{sec:deprojection}, we require SZ effect images that resolve angular scales comparable to $r_{2500}$, defined as the radius within which the mean enclosed density is 2500 times the critical density at the cluster's redshift.\footnote{We later refer to $r_{500}$, defined analogously. For this cluster sample, $r_{500}\approx2.2\,r_{2500}$.} Based on the publicly available data described in Section~\ref{sec:SZ_data}, this reduces the sample size to 15. However, the SZ data for one of these clusters (RX~J1524.2$-$3154) is significantly contaminated by nearby galactic dust emission (see Section~\ref{sec:astrophysical uncertainties}), necessitating its removal from our study and resulting in a final sample of 14.
The cluster sample and available data are summarized in Table~\ref{tab:data}.

\begin{table*}
	\centering
	\caption{
    Clusters in our data set, and basic parameters of the X-ray and SZ data.
    Column [1] name; 
    [2], [3] J2000 coordinates of the cluster center, from \citet{Mantz1402.6212};
    [4] clean {\it Chandra} exposure time;
    [5] signal-to-noise ratio of the Bolocam SZ effect detection, from  \citet{Sayers2016};
    [6] signal-to-noise ratio of the {\it Planck} SZ effect detection, from the \citet{Planck2016_XXII} and \citet{Sayers2016}.
    }
	\label{tab:data}
	\begin{tabular}{lccccc} 
		\hline
		\hline
		Cluster & RA & Dec & Chandra exp. & Bolocam S/N & Planck S/N \\
		& & & (ks) &  & \\
		\hline
		Abell~2029           &  15:10:55.9  &  +05:44:41.2    & 118.9             & ---              & 23.2             \\
Abell~478            &  04:13:25.2  &  +10:27:58.6    & 129.4             & ---              & 15.8             \\
PKS~0745$-$191       &  07:47:31.7  &  $-$19:17:45.0  & 148.8             & ---              & 21.3             \\
Abell~2204           &  16:32:47.1  &  +05:34:31.4    & $\phantom{0}$90.1 & 22.3             & 16.3             \\
RX~J2129.6+0005      &  21:29:39.9  &  +00:05:18.3    & $\phantom{0}$36.7 & $\phantom{0}$8.0 & $\phantom{0}$4.8 \\
Abell~1835           &  14:01:02.0  &  +02:52:39.0    & 183.6             & 15.7             & 14.4             \\
MS~2137.3$-$2353     &  21:40:15.2  &  $-$23:39:40.0  & $\phantom{0}$50.9 & $\phantom{0}$6.5 & $\phantom{0}$2.9 \\
MACS~J1931.8$-$2634  &  19:31:49.6  &  $-$26:34:32.7  & 104.0             & 10.1             & $\phantom{0}$6.1 \\
MACS~J1115.8+0129    &  11:15:51.9  &  +01:29:54.3    & $\phantom{0}$44.3 & 10.9             & $\phantom{0}$7.1 \\
MACS~J1532.8+3021    &  15:32:53.8  &  +30:20:58.9    & 102.4             & $\phantom{0}$8.0 & $\phantom{0}$1.8 \\
MACS~J1720.2+3536    &  17:20:16.8  &  +35:36:27.0    & $\phantom{0}$51.7 & 10.6             & $\phantom{0}$6.5 \\
MACS~J0429.6$-$0253  &  04:29:36.1  &  $-$02:53:07.5  & $\phantom{0}$19.3 & $\phantom{0}$8.9 & $\phantom{0}$4.1 \\
RX~J1347.5$-$1145    &  13:47:30.6  &  $-$11:45:10.0  & 206.5             & 36.6             & 11.2             \\
MACS~J1423.8+2404    &  14:23:47.9  &  +24:04:42.3    & 123.7             & $\phantom{0}$9.4 & $\phantom{0}$1.8 \\

		\hline
	\end{tabular}
\end{table*}

\subsection{\textit{Chandra} X-ray data} \label{sec:xdata}

Our procedure for reducing and cleaning the {\it Chandra} data is described in detail by \citet{Mantz1402.6212, Mantz1502.06020}.
We depart from this only by using a more recent version of the {\it Chandra} analysis software and calibration files (respectively, versions 4.9 and 4.7.4).
A direct comparison with {\sc ciao} 4.6.1/{\sc caldb} 4.6.2, as employed by \cite{Mantz1509.01322}, reveals negligible changes in the derived gas densities and temperatures.

Our methods for determining deprojected (3-dimensional) density and temperature profiles from the X-ray data are also described in earlier work \citep{Mantz1402.6212, Mantz1509.01322}.
In brief, using the {\sc projct} model in {\sc xspec},\footnote{\url{https://heasarc.gsfc.nasa.gov/docs/xanadu/xspec/}} the ICM is modeled as a series of concentric, spherical and (individually) isothermal shells.
No assumptions about the underlying mass profile are required.
The gas density, temperature and metallicity as a function of radius are fitted simultaneously, with covariances among the different quantities at different radii fully accounted for.

The centers of the X-ray thermodynamic profiles are chosen to reflect symmetry at radii near $r_{2500}$, typically $\sim500$\,kpc for this sample \citep{Mantz1402.6212}, rather than at small scales.
As a result, density determinations near the cluster centers may be inaccurate simply because the profile center does not sit on the peak of the X-ray brightness.
In addition, genuine small-scale variations in density and temperature due to, e.g., AGN feedback, are commonly seen within the central few tens of kpc in even the most relaxed clusters.
We therefore inflate the uncertainties for pressures at the small radii where such disturbances are observed, as identified by \citet{Mantz1402.6212}, to 50 per cent of the nominal measured values.
We verified that our joint X-ray/SZ profile fits (Section~\ref{sec:deprojection}) are insensitive to this choice (e.g., inflating the uncertainties to 25 per cent instead of 50 per cent changes the best-fitting X-ray/SZ pressure ratio by only $\sim 0.5$ per cent), as one might expect, given that the impacted scales are unresolved in the SZ data.

\subsection{Sunyaev-Zel'dovich effect data}
\label{sec:SZ_data}

We use two separate publicly available SZ effect data sets for this analysis: the R2.00 all-sky {\it Planck} $y$ map produced using the MILCA algorithm \citep{Planck2016_XXII}\footnote{\url{https://irsa.ipac.caltech.edu/data/Planck/release_2/all-sky-maps/ysz_index.html}} and the "filtered\_image" targeted Bolocam maps \citep{Sayers2013}.\footnote{\url{https://irsa.ipac.caltech.edu/data/Planck/release_2/ancillary-data/bolocam/bolocam.html}} As demonstrated by \citet{Sayers2016}, these two data sets provide a consistent measurement of the SZ effect signal within the calibration uncertainties, and they can thus be combined to probe a wide range of angular scales. The {\it Planck} $y$ map has an effective point spread function (PSF) with a full width at half maximum (FHWM) of 10\arcmin, a flux calibration uncertainty of better than 0.1 per cent in the most sensitive SZ effect bands at 100 and 143\,GHz \citep{Planck2016_VIII}, and fidelity on the SZ effect signal on all angular scales. The Bolocam maps have a size of 14\arcmin\ square, a PSF with a FWHM of 58\arcsec, and high-pass filtering that limits their fidelity on angular scales larger than $\sim 10$\arcmin. As found by \citet{Sayers2019}, the flux calibration uncertainty when referenced to the planetary model of the \citet{Planck2017_LII} is 1.6 per cent, due to a combination of measurement uncertainty in the calibration observations, variations in the atmospheric transmission, and variations in the PSF. To good approximation, this 1.6 per cent uncertainty should be random between clusters. Furthermore, it is likely a slight overestimate of the true variation in our analysis, since some of the PSF variation noted by \citet{Sayers2019} is due to differences in spectral response to the thermal and kinematic SZ effects, while this work is focused solely on the thermal SZ effect.

While the SZ effect data have been processed to remove and/or account for all of the relevant sources of astrophysical contamination, we find some spurious signals in the {\it Planck} $y$ maps. This is particularly true for low-redshift clusters that subtend large angular sizes, and those located in regions of significant galactic dust emission. 
To mitigate the impact of these unwanted signals, we remove some regions of the $y$ maps from our study. First, we search for any pixels in the radial range $2 \le r/r_{500} \le 6$ with a signal-to-noise ratio (S/N) $\ge 5$. 
Within this radial range, the SZ effect signal falls below this threshold, so any such pixels are due to spurious signals.
Such flagging occurs for the three lowest-redshift clusters in our sample (Abell~2029, Abell~478, and PKS~0745$-$191). All pixels inside that radial range and within 2\,FWHM (20\arcmin) of these high S/N pixels are then removed. In addition, we search for any known clusters from the MCXC catalog \citep{Piffaretti2011} within a projected distance of 6\,$r_{500}$ from the target cluster. Two such interlopers are identified; MCXC\,J1511.3+0619 is $\approx37\arcmin$ NE of Abell~2029 and MCXC\,J2139.0$-$2333 is $\approx17\arcmin$ NW of MS~2137.3$-$2353. All pixels within circular apertures centered on the positions of the interlopers were removed. Given the locations of these objects, and the desire to retain the SZ signal from the cluster of interest, a 20\arcmin\ diameter aperture was used for the Abell~2029 image, and a 10\arcmin\ diameter aperture for the MS~2137.3$-$2353 image.

\section{Methods} \label{sec:methods}

In this section, we detail the procedure we followed to derive the value of $H_0$ with the combination of X-ray and SZ observations.

\subsection{Mitigating observer bias}

As noted in the introduction, the precise value of the Hubble Constant is currently the subject of some controversy.
While our estimate of the systematic uncertainty indicates that we will not have the precision to distinguish between the values preferred by CMB and distance ladder data, we nevertheless felt it was a useful exercise to take steps to prevent observer bias from influencing our results.
This is straightforwardly accomplished, given that $H_0$ appears strictly as a multiplicative term in the model fitted to the data (Section \ref{sec:cosmodel}).

To be precise, our analysis proceeded in 3 phases.
In the first phase, before being input to the joint X-ray and SZ fitting procedure (Section \ref{sec:deprojection}), the X-ray pressure data for all clusters were multiplied by a single random and unknown factor obtained from a uniform distribution between 0.5 and 1.5 (i.e., analogous to an overall shift in calibration). In addition, the X-ray pressure data for each cluster were multiplied by different random and unknown factors obtained from a normal distribution with a mean of 1 and a standard deviation itself obtained from a uniform distribution between 0.1 and 0.5 (i.e., analogous to an intrinsic cluster-to-cluster scatter with an unknown magnitude).
This allowed the joint fitting procedure to be vetted and decisions about the modeling of measurement uncertainties to be made without revealing, for example, which clusters were likely to be outliers in their X-ray/SZ pressure ratios.
We also converged on expectations from the literature for various potential systematic effects during this phase, while the true cluster-to-cluster scatter was still hidden.
In the second phase, the individual unknown factors were removed, but the overall scaling remained in place.
The average X-ray/SZ ratio thus remained hidden, while possible systematic trends of this ratio with, for example, dust emission, were investigated (Section~\ref{sec:astrophysical uncertainties}).
Only after finalizing decisions about the treatment of systematics and the associated priors on nuisance parameters, and testing the performance of our likelihood function (Section~\ref{sec:model}) on mock data, did we proceed with the third phase of analysis with the input data unaltered. No changes to the analysis methods were implemented during this third phase; therefore all of the decisions made
while the X-ray/SZ ratio was hidden were retained in determining our final results.

\subsection{Joint deprojection of pressure profiles}
\label{sec:deprojection}

We assumed spherical symmetry and fit a single pressure profile model to the X-ray and SZ data for a single cluster simultaneously. Our fitting algorithm is based on the technique developed by \citet{Sayers2016}, which we briefly summarize. The radial pressure profile is assumed to have a set of values ${\mysub{P}{e}}_{,i}$ at discrete, logarithmically spaced radii $R_i$. Between $R_i$, the pressure is modeled as a power law with a constant exponent. The pressure is assumed to be zero beyond $6\,r_{500}$. For comparison to the SZ effect maps, the model was projected and convolved with the appropriate PSF shape. In the case of Bolocam, the effective high-pass filter of the data processing was also applied. Furthermore, radially dependent relativistic corrections to the SZ effect signal were computed, based on the deprojected temperature profile obtained from the X-ray analysis, using {\sc SZpack} \citep{Chluba2012,Chluba2013}. Following the empirically demonstrated results of \citet{Sayers2016}, we assumed an effective band center of 143~GHz when computing the relativistic corrections to the {\it Planck} $y$ maps. For the {\it Chandra} X-ray data, the model was directly compared with the deprojected pressures obtained from that analysis. To allow for a difference in the X-ray-derived and SZ effect-derived pressures, which is what we aimed to measure in order to constrain our cosmological model (Section~\ref{sec:cosmodel}), the values of ${\mysub{P}{e}}_{,i}$ were multiplied by a constant factor, \rxs, prior to comparison with the deprojected X-ray profile.

\begin{figure*}
	\includegraphics[width=\textwidth]{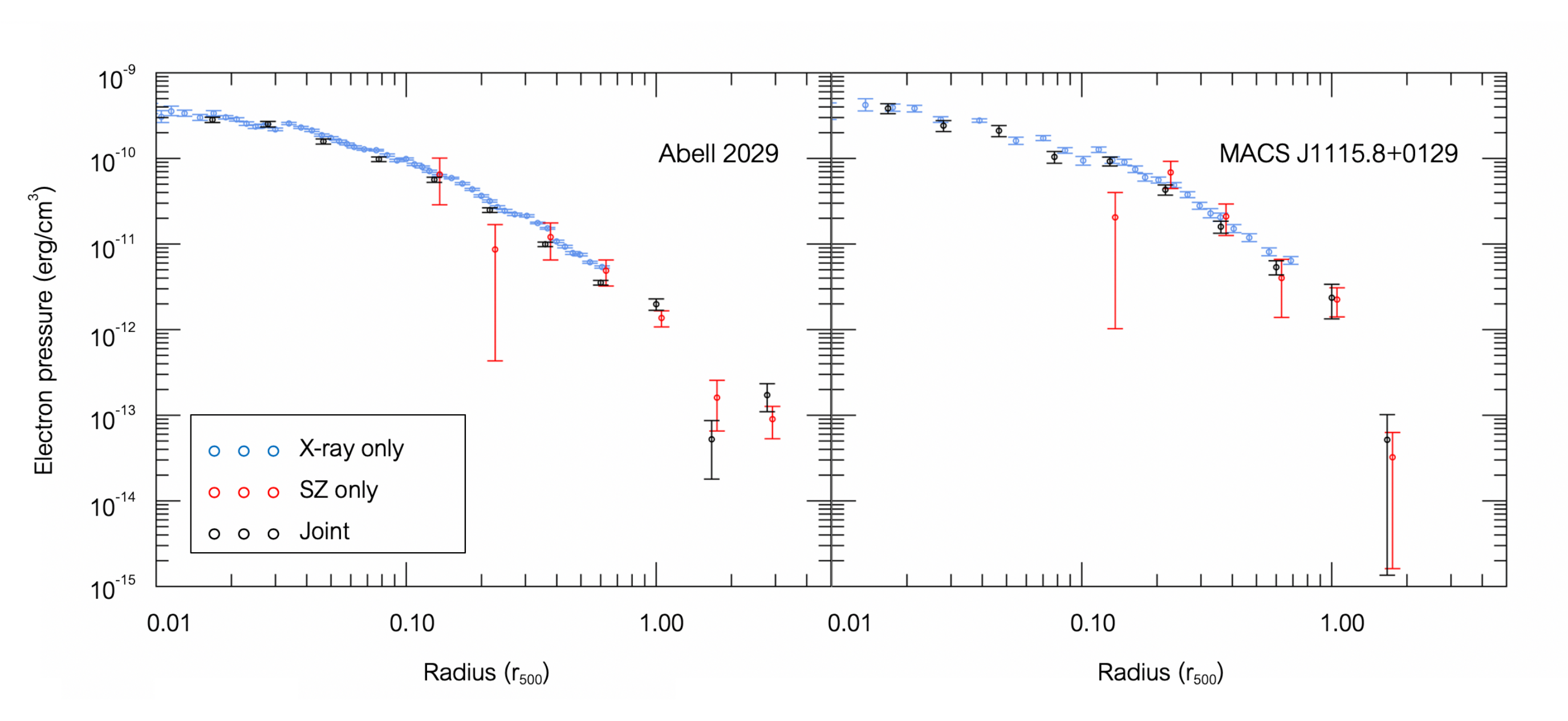}
	\caption{
	Deprojected electron pressure profiles of Abell~2029 (left) and MACS\,J1115.8+0129 (right) comparing fits using only X-ray data, only SZ data, and both X-ray and SZ data. 
	The fitted value of \rxs{} in general results in a slight multiplicative offset between the X-ray only profiles and the joint profiles.
	The data were normalized assuming a reference cosmology of $H_0=70\,\kmsMpc$, $\Omegam = 0.3$, and $\Omega_{\Lambda} = 0.7$ (this choice has no impact on the cosmological results obtained; see Section~\ref{sec:cosmodel}).
	}
    \label{fig:pressure profiles}
\end{figure*}

One advantage of using this method to constrain \rxs\ is that the fit naturally up-weights the angular scales probed by both the X-ray and SZ effect data while down-weighting other angular scales. Specifically, at small radii unresolved by the PSF of the SZ effect maps, the values of ${\mysub{P}{e}}_{,i}$ are constrained almost entirely by the X-ray data, while at large radii beyond the extent of the X-ray deprojections, the values of ${\mysub{P}{e}}_{,i}$ are determined exclusively by the SZ effect data. Therefore, the value of \rxs\ is determined almost entirely from the intermediate radii probed by both observables, corresponding to approximately 0.1--1.0\,$r_{500}$ (see Figure~\ref{fig:pressure profiles}). This radial range is largely outside of the influence of the central AGN \citep[e.g.,][]{McNamara2007}, and interior to the regions strongly impacted by active accretion \citep[e.g.,][]{Lau2015}. As a result, it is also the region where the pressure profiles demonstrate the least amount of intrinsic cluster-to-cluster scatter \citep[e.g.,][]{Arnaud2010,Sayers2013,Planelles2017}, and the region most likely to be free from biases due to gas clumping \citep[e.g.,][]{Urban2014,Eckert2015,Battaglia2015,Planelles2017}. Therefore, while this radial range is dictated by the properties of the observational data, it is also close to optimal for our study based on physical considerations.

It is not possible to accurately determine the small amount of correlated pixel-to-pixel noise in the Bolocam maps from either the data themselves or from models of the instrument and atmospheric noise fluctuations. We therefore followed the technique described by \citet{Sayers2016}, and assumed that the map-space noise is described by a diagonal covariance matrix in order to determine the best-fitting values of ${\mysub{P}{e}}_{,i}$ and \rxs. For simplicity, we treated the {\it Planck} and X-ray data in an analogous manner, neglecting correlations and any other non-idealities in those measurements. To determine the best-fitting values of ${\mysub{P}{e}}_{,i}$ and \rxs, we used a generalized least squares (GLS) fitting algorithm \citep{Markwardt2009}.

Because the diagonal noise covariance matrices are an imperfect description of the data, we again followed the technique of \cite{Sayers2016} to quantify both the biases and the uncertainties on the fitted parameter values. Random realizations of Bolocam noise maps are included as part of the public data release, and the best-fitting pressure model was added to 1000 such realizations. An analogous procedure was used for {\it Planck}, based on noise realizations generated from both the homogeneous and inhomogeneous noise spectra included in the public data release. For the X-ray data, 1000 random samples from the Markov chain Monte Carlo analysis originally used to constrain the profiles were used. Each of these 1000 data realizations was then fit using the same (diagonal) noise covariance matrix as the real data. In the case of the Bolocam data, a Gaussian random flux scaling with a root mean square (rms) of 1.6 per cent was applied to each noise realization (see Section~\ref{sec:SZ_data}).

\begin{figure*}
	\includegraphics[width=\textwidth]{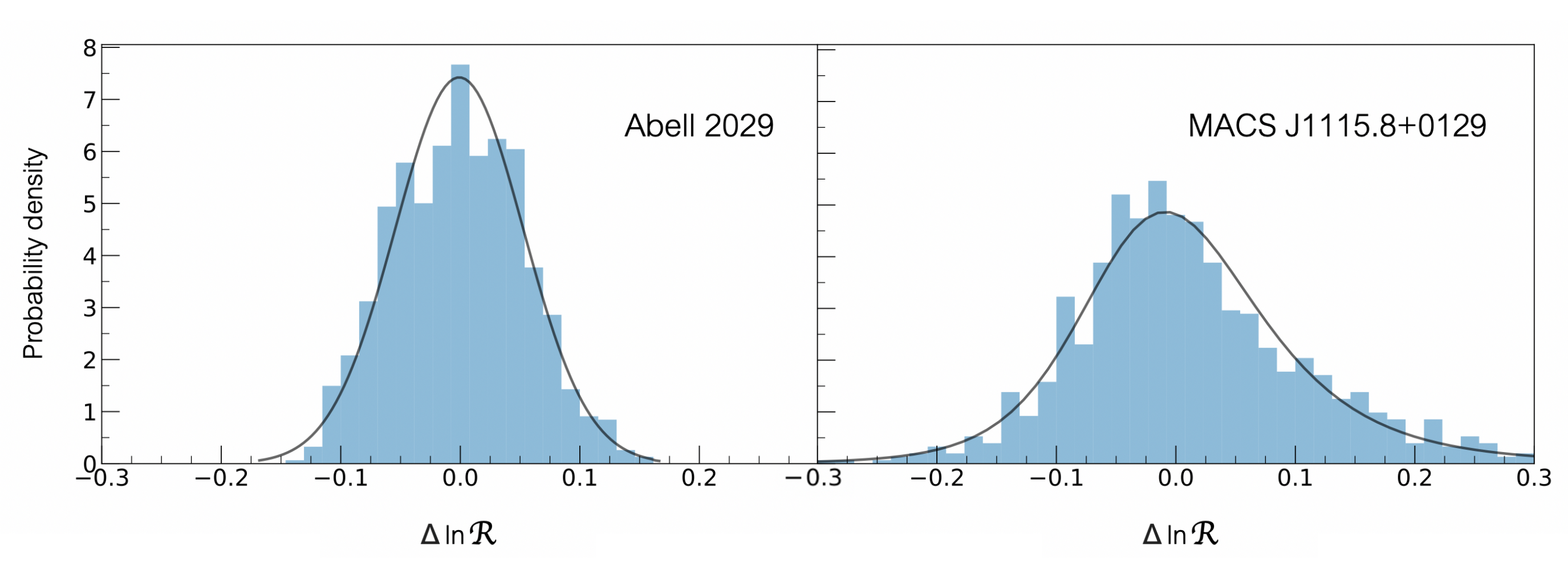}
	\caption{
	Normalized distributions of measured $\ln\rxs{}$ values from 1000 realizations of the observational noise, relative to the median, for two example clusters.
	Curves show the best-fitting non-central $t$ distributions.
    }
    \label{fig:nct dist}
\end{figure*}

Empirically, the distribution of best-fitting values of \rxs\ from these data realizations often displayed both significant skewness and heavier-than-Gaussian tails. We therefore described the distributions of $\ln\rxs{}$ for each cluster using the more flexible non-central $t$ distribution (Appendix~\ref{sec:nct}; see Fig. \ref{fig:nct dist} for an example).
Table~\ref{tab:fitresults} shows the best-fitting parameter values for each cluster.
A Kolmogorov-Smirnov test verified that this model provides an accurate description in each case. In addition, for each cluster, we compared the median value obtained from the 1000 data realizations to the best-fitting value of \rxs\ obtained from the real data. We found no outliers, and an average difference of $\Delta \ln\rxs{} = 0.024 \pm 0.026$. This bias, which is likely a result of the diagonal noise covariance approximation used in the fits, was assumed to be identical for every cluster and is accounted for as an overall systematic effect in our final analysis (Section~\ref{sec:fit_bias}).

\begin{table*} 
	\centering
	\caption{Results of the joint fit to the X-ray and SZ data for our sample of clusters, along with redshifts. Column [1] name;
	[2] redshift;
	[3] best-fitting \rxs{} values and standard deviations based on 1000 noise realizations (these provide a simple estimate of the precision, while the measurement error distribution is more accurately given by the next columns);
	[4--7] non-central $t$ distribution parameters describing the sampling distribution of $\ln\rxs{}$ (see Appendix~\ref{sec:nct});
	[8] the linear slope relating the estimate of \rxs{} with the temperature calibration bias (see Section~\ref{sec:xraycal}); [9] the angular diameter distance (in Mpc) estimated from the \rxs{} value of each cluster as a function of redshift, assuming a flat $\Lambda$CDM cosmology with $\Omegam=0.3$.
	}
	\label{tab:fitresults}
	\begin{tabular}{lccrrrrrr}
		\hline
		\hline
		Cluster & Redshift & \rxs{} & \multicolumn{1}{c}{$\nu$} & \multicolumn{1}{c}{$c$} & \multicolumn{1}{c}{$\mu$} & \multicolumn{1}{c}{$\sigma$} & \multicolumn{1}{c}{$\alpha$} & \multicolumn{1}{c}{$d_A(z)$ (Mpc)} \\
		\hline
        Abell~2029          & 0.078 & 1.264 $\pm$ 0.073 & 340.00 &    0.066 & $-$0.0044 & 0.054 & 0.897 & $237.5^{+17.2}_{-16.0}$ \vspace{1ex}\\
Abell~478           & 0.088 & 1.229 $\pm$ 0.086 & 340.00 &    0.156 & $-$0.0060 & 0.066 & 0.861 & $278.3^{+22.2}_{-20.6}$ \vspace{1ex}\\
PKS~0745$-$191      & 0.103 & 1.400 $\pm$ 0.106 &  16.80 & $-$3.873 &    0.2276 & 0.059 & 0.933 & $249.8^{+37.9}_{-32.9}$ \vspace{1ex}\\
Abell~2204          & 0.152 & 0.909 $\pm$ 0.051 &   4.16 &    1.056 & $-$0.0410 & 0.037 & 0.850 & $815.1^{+57.1}_{-53.4}$ \vspace{1ex}\\
RX~J2129.6+0005     & 0.235 & 1.148 $\pm$ 0.128 &   3.94 &    0.825 & $-$0.0617 & 0.076 & 0.988 & $740.4^{+93.6}_{-83.1}$ \vspace{1ex}\\
Abell~1835          & 0.252 & 1.071 $\pm$ 0.090 &   4.09 &    0.338 & $-$0.0208 & 0.059 & 0.934 & $887.6^{+142.0}_{-122.4}$ \vspace{1ex}\\
MS~2137.3$-$2353    & 0.313 & 1.042 $\pm$ 0.112 &   2.06 &    0.256 & $-$0.0157 & 0.053 & 0.789 & $1064.8^{+223.8}_{-185.0}$ \vspace{1ex}\\
MACS~J1931.8$-$2634 & 0.352 & 0.925 $\pm$ 0.093 &   3.82 &    0.574 & $-$0.0357 & 0.066 & 0.855 & $1477.1^{+415.4}_{-324.2}$ \vspace{1ex}\\
MACS~J1115.8+0129   & 0.355 & 1.225 $\pm$ 0.135 &   4.48 &    0.748 & $-$0.0580 & 0.076 & 0.932 & $859.1^{+205.0}_{-165.5}$ \vspace{1ex}\\
MACS~J1532.8+3021   & 0.363 & 1.300 $\pm$ 0.126 &   4.04 &    0.449 & $-$0.0251 & 0.072 & 0.928 & $773.3^{+195.0}_{-155.7}$ \vspace{1ex}\\
MACS~J1720.2+3536   & 0.391 & 1.141 $\pm$ 0.215 &   9.04 &    0.365 & $-$0.0395 & 0.151 & 0.890 & $1044.9^{+349.5}_{-261.9}$ \vspace{1ex}\\
MACS~J0429.6$-$0253 & 0.399 & 0.940 $\pm$ 0.167 &   4.94 &    0.451 & $-$0.0565 & 0.124 & 0.964 & $1579.6^{+768.8}_{-517.1}$ \vspace{1ex}\\
RX~J1347.5$-$1145   & 0.450 & 1.132 $\pm$ 0.065 &  13.22 &    0.319 & $-$0.0147 & 0.048 & 0.763 & $1127.7^{+273.8}_{-220.3}$ \vspace{1ex}\\
MACS~J1423.8+2404   & 0.543 & 1.105 $\pm$ 0.116 &   3.39 &    1.153 & $-$0.0691 & 0.059 & 0.719 & $1296.8^{+724.7}_{-464.9}$ \vspace{1ex}\\

		\hline
	\end{tabular}
\end{table*}

\subsection{Cosmological Model} \label{sec:cosmodel}

The X-ray and SZ signals predicted for a given cluster pressure model have different dependences on the cosmic distance to the cluster's redshift; this is the origin of the cosmological sensitivity of $\rxs$.
In the case of X-ray data, measurements of temperature depend on the shape of the observed spectrum, and have no dependence on the distance to the source.
However, the X-ray density determination follows from relating the observed flux to the ICM emissivity, and thus involves factors of the luminosity distance, $\dL(z)$, and (via the volume element) the angular diameter distance, $\dA(z)$.
In detail, the density of electrons and protons, $\nelec$ and $\nH$, inferred from X-ray data depend on cosmic distances as $\nelec\nH \propto \dL(z)^2/\dA(z)^3$ (see, e.g., discussion in \citealt{Mantz1402.6212}).
We assume a canonical ratio of $\nelec/\nH=1.2$ throughout, though we note that, in principle, this factor has a small sensitivity to the primordial helium abundance.
Thus, the physical electron pressure inferred for a cluster at redshift $z$, given the X-ray data, follows the relation $\nelec kT \propto \dL(z)/\dA(z)^{3/2}$.
The projected Compton $y$ parameter is straightforwardly proportional to a line-of-sight integral of the electron pressure, where the integration element includes a factor of the angular diameter distance.
Hence, the pressure we would infer from the SZ data is proportional to $\dA(z)^{-1}$.
Combining this with the X-ray pressure dependence, the inferred ratio of X-ray to SZ pressure, \rxs, is proportional to $\dL(z)/\dA(z)^{1/2}$.
This is conventionally simplified to $\rxs \propto \dA(z)^{1/2}$, using the distance-duality relation.

In practice, both the X-ray pressure profiles and the joint fitting were performed assuming distances given by a reference flat \LCDM{} model with $\Omegam=0.3$ and $H_0=70\,\kmsMpc$.
In our analysis, we will therefore need to predict the values of \rxs{} that would be inferred under that assumption, given some possibly different trial cosmology.
In this case, we have that the predicted $\rxs \propto \sqrt{\dref(z)/\dA(z)}$, where $\dref(z)$ is the distance to a cluster's redshift according to the reference model, and $\dA(z)$ is the distance in the trial model.
This is the dependence that appears in the complete likelihood for our data (Section~\ref{sec:model}).
Note that the proportionality above captures exactly the impact of this assumed reference model, such that our cosmological results are not sensitive to it.
We have verified explicitly that we obtain statistically identical results when the entire analysis is repeated using a different reference cosmology.

\subsection{Potential systematics} \label{sec:systematics}

Our selection of dynamically relaxed clusters, and the radial range to which the data are sensitive, minimizes a number of the systematic effects discussed in the literature for similar analyses \citep[e.g.,][]{Bonamente2006,Kozmanyan2019}. In this section, we describe various potential systematic effects and their expected impacts.

We can consider two distinct effects on our results: cluster-to-cluster scatter in the value of $H_0$ inferred from a single system, and bias in the average value of $H_0$ from clusters in the sample.
The former is addressed by including an intrinsic scatter term in the model and fitting for it simultaneously with other parameters.
The latter, an overall bias, is thus of primary concern in the sections below.
In general, our approach is to account for potential sources of systematic bias and their uncertainty when they can be rigorously quantified, even if they are not large enough to impact the overall error budget.

\begin{figure*}
	\includegraphics[width=\textwidth]{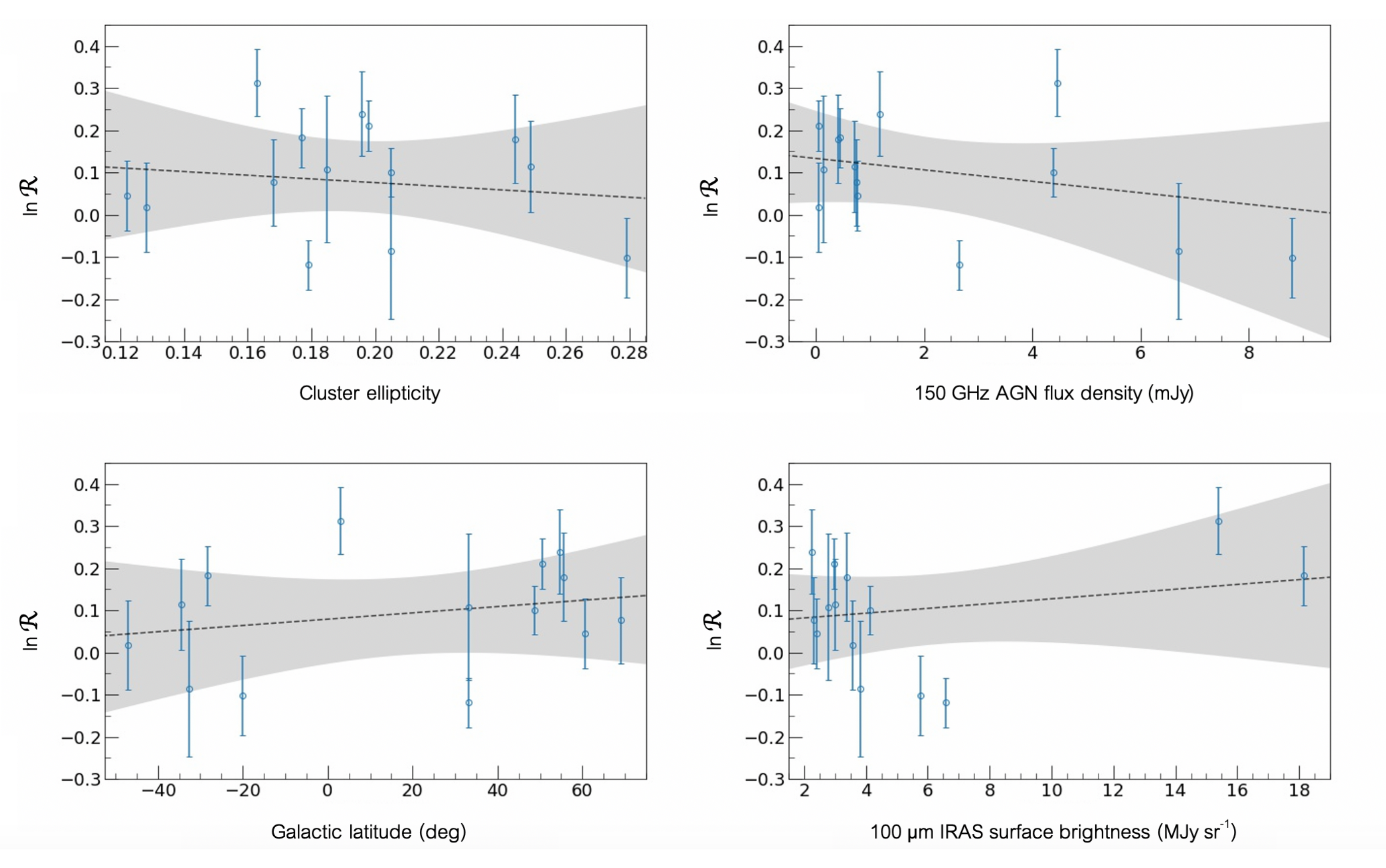}
	\caption{Natural log of the X-ray/SZ pressure ratio as a function of observable proxies for various potential systematics that might be expected to impact the measurement. Shaded regions show the uncertainties on a linear fit to the data, accounting for intrinsic scatter. Formally, each of these trends is consistent with a constant.
	The plotted values are provided in Tables~\ref{tab:fitresults} and \ref{tab:fig3 data}.
	}
    \label{fig:scaling_factors}
\end{figure*}

\subsubsection{Astrophysical uncertainties}
\label{sec:astrophysical uncertainties}

\paragraph{Asphericity:}
\label{sec:asphericity}
Our deprojection analysis assumes spherical symmetry, while the bulk geometry of clusters is better described by a triaxial ellipse \citep{Jing2002, Bonamigo2015, Lau2020}. This is expected to result in a non-zero cluster-to-cluster scatter in the measured value of \rxs, due to the differing density dependences of the X-ray and SZ signals. Quantitatively, the resultant scatter in $H_0$ inferred from a given cluster is expected to be $\simeq 10$--15 per cent, with a potential bias in the average value of $H_0$ derived from a large sample of randomly oriented clusters as large as $\sim 3$ per cent due to asymmetries in this scatter \citep{Sulkanen1999,Kawahara2008,Kozmanyan2019}. 
However, for a sample of rounder clusters, corresponding to 3-D axial ratios of 0.9:0.9:1.0, \citet{Kawahara2008} found that the scatter and bias can be lower than these nominal values;
while our selection is not biased towards rounder clusters in projection, it does provide a more spherical sample than the complete population.
Interestingly, while we might still expect to see some trend in the measured \rxs{} values with projected cluster ellipticity (as measured by \citealt{Mantz1502.06020}), the data are consistent with a constant value (top-left panel of Figure~\ref{fig:scaling_factors}).
This suggests that any residual variance in \rxs{} is not due to asphericity on the intermediate scales probed by these particular ellipticity measurements (typically $\sim0.5$--$1\,r_{2500}$), consistent with the expectation that the cluster gas is more spherical at $\sim r_{2500}$ than at larger scales ($r\gtsim r_{500}$). The model fit in Section~\ref{sec:model} includes a free parameter accounting for residual intrinsic scatter, irrespective of its origin.

\paragraph{Gas clumping:}
Realistic gas distributions that include clumping, which enhances the X-ray emission, can result in scatter and biases in the value of \rxs\ comparable in magnitude to those produced by asphericity \citep{Kawahara2008, Kozmanyan2019}. As noted above, our fits are constrained almost entirely from data probing the radial range 0.1--1.0\,$r_{500}$, where such clumping is expected to be at a minimum from both simulations \citep[e.g.,][]{Battaglia2015, Planelles2017} and observations \citep[e.g.,][]{Simionescu2011, Urban2014, Eckert2015}. 
For these radii, simulations place the expected bias due to clumping at $\sim3$--$5$ per cent in \rxs{} \citep{Roncarelli2013, Ansarifard2020}.
Observational limits on the combined impact of clumping and turbulent density fluctuations in comparable regions of the Coma cluster are at the $\sim4\pm1$ per cent level at scales of tens of kpc \citep{Zhuravleva1906.06346}; since Coma is a merging system, this suggests that clumping in relaxed clusters may be somewhat smaller than the predictions from simulations.
A potential bias at this level is small compared with other systematic
effects, and therefore has not been explicitly included in our analysis.

\paragraph{Helium sedimentation:}
The relationship between electron density and X-ray emissivity depends on the chemical composition of the gas, with hydrogen and helium comprising the overwhelming majority of nucleons present.
In principle, over cosmological timescales, the heavier helium nuclei may sink within the cluster potential, with the effects on density determination being strongest in the inner regions (e.g.\ \citealt{Ettori0603383}, and references therein).
However, this process can be suppressed substantially in the presence of magnetic fields, especially when temperature gradients cause the fields to become arranged azimuthally \citep{Parrish0905.4500}.
Mixing, driven by bulk and turbulent motions in the ICM, sourced by galaxy motions and merger activity and AGN feedback \citep{McNamara2007,Fabian1204.4114,Zhuravleva1410.6485,Hitomi1607.04487,Lau1705.06280}, will also act to suppress sedimentation.
While transport processes within the ICM remain a subject of active research (e.g.\ \citealt{Berlok2007.00018,Drake2007.07931}; and references therein), given the magnetic field strengths of tens of $\mu$G typically observed in relaxed cluster cores \citep{Carilli0110655} and the down-weighting of the smallest radii in our analysis, we expect the effects of helium sedimentation on the measurements to be minimal.

\paragraph{Galactic absorption:}
Our X-ray analysis assumes Galactic equivalent hydrogen absorbing column densities given by the LAB \ion{H}{i} survey \citep{Kalberla2005} when those values are $\NH<10^{21}$\,cm$^{-2}$.
For larger column densities, the values based on only \ion{H}{i} measurements are known to be inaccurate, and we instead make $\NH$ as a free parameter in the fit.
For Abell\,478 specifically, the column density varies spatially over the cluster field, and we additionally restrict the energy range of our analysis to minimize the impact of this effect (see \citealt{Mantz1402.6212} for details).
With these measures, we expect any residual systematics related to Galactic absorption to be negligible.

\paragraph{X-ray foregrounds/backgrounds:}
Our analysis models non-cluster X-ray signal as a combination of emission from unresolved AGN, activations due to non-X-ray particles hitting the detector, and foreground emission from the Galactic halo and local hot bubble. The background components are accounted for in the standard way via rescaled ``blank-sky'' observations, while the foregrounds are fitted simultaneously with the cluster model when statistically required by the data (see \citealt{Mantz1402.6212} for details). Any systematic inadequacies in this procedure may be empirically considered contributors to the ``X-ray calibration'' uncertainty (below).

\paragraph{Radio galaxies:} \label{sec:radiogalaxies}
The central cluster galaxy often hosts an AGN with bright synchrotron emission that can contaminate the SZ effect signal. {\it Planck}'s multi-frequency coverage efficiently removes this contamination, and a combination of measurements at 1.4 and 30\,GHz were used to model and subtract this signal from the Bolocam images \citep[see][]{sayers2013radio}. Uncertainties in the removal of these sources are included in the noise realizations used when fitting the pressure profiles, and are generally negligible in the overall error budget. To further search for any residual AGN contamination in the SZ effect data, we examined the recovered values of \rxs\ as a function of AGN flux density near 150 GHz \citep[][see the top-right panel of Fig.~\ref{fig:scaling_factors}]{sayers2013radio}. 
For clusters in the sample with both \textit{Planck} and Bolocam data, this AGN flux density was the predicted flux density near 150 GHz extrapolated from lower-frequency measurements by \cite{sayers2013radio}. For the three clusters without Bolocam data (Abell~2029, Abell~478 and PKS~0745$-$191), this flux density was estimated by fitting a power law to photometric measurements available in the literature between 1 GHz and $\sim$25 GHz 
\citep{PKS0745photometry3,PKS0745photometry1,A2029photometry4,A2029photometry5,PKS0745photometry2,A2029photometry6,A2029photometry1,A2029photometry8,A478photometry1,A2029photometry7,A2029photometry2}. We found no correlation between the value of \rxs\ and the flux density of the central AGN, indicating there are no significant un-modeled biases. 
Furthermore, as found by \citet{sayers2013radio}, contamination from non-central AGN and unassociated field radio galaxies is much less common, and has been removed from our data according to the measurements in that work.

\paragraph{Kinematic SZ effect:}
The radial motion of the clusters with respect to the rest frame of the CMB imparts a Doppler shift to the scattered photons that results in the signal known as the kinematic SZ effect \citep{Sunyaev1972,Sunyaev1980}. We account for the variations in the measured SZ effect signal due to this Doppler shift as part of the noise realizations fitted in our deprojection analysis. Specifically, we follow the approach of \citet{Mueller2015}, assuming a 300\,km\,s$^{-1}$ rms scatter in the line-of-sight velocity of each cluster based on the simulations of \citet{Sheth2001}. The kinematic SZ effect signal resulting from random velocities drawn from a Gaussian distribution with this rms scatter is added to each noise realization of the Bolocam data. Because the {\it Planck} $y$ map is constructed in a manner that removes the kinematic SZ effect signal, we do not add any such signal to those noise realizations. Given the ICM temperatures of the clusters in our sample, this velocity scatter corresponds to approximately 5 per cent of the total thermal SZ signal, comparable to the typical measurement noise. We find, however, that this has a negligible impact on the derived values of \rxs, likely because the overall signal level is largely constrained by the {\it Planck} data, which have sufficient spectral coverage to distinguish the kinematic and thermal SZ effects.

\paragraph{Primary CMB anisotropies:}
The primary CMB anisotropy signal is largely removed from the {\it Planck} $y$ map. For Bolocam, random realizations of these anisotropies are included in the noise realizations (e.g., see \citealt{Sayers2013}). The resulting parameter uncertainties due to the CMB anisotropies are generally small, but non-negligible, compared with those due to instrument noise.

\paragraph{Cosmic infrared background:}
Analogous to primary CMB anisotropies, fluctuations due to emission from the cosmic infrared background are included in the Bolocam noise realizations (see \citealt{Sayers2013}) and removed from the {\it Planck} $y$ map.

\paragraph{Diffuse radio emission:}
The typical spectrum of diffuse radio emission renders it negligible at the observing frequencies relevant to {\it Planck} and Bolocam (e.g., \citealt{vanWeeren2019}).

\paragraph{Galactic dust emission:}
While the {\it Planck} algorithm used to create the $y$ maps largely removes contamination from galactic dust emission, a non-negligible residual can remain. As detailed in Section~\ref{sec:SZ_data}, we identified regions of the map with likely contamination and removed them from our analysis. This was based on a search for pixels with a S/N~$\ge 5$, excluding the clusters' central regions. To characterize the potential impact of residual dust emission that was not removed via this procedure, we searched for trends between the measured values of \rxs\ and the expected level of overall dust contamination towards each cluster. As a proxy for the latter, we considered both the galactic latitude and the 100\,\micron{} IRAS surface brightness towards the center of each cluster
\citep[][see the bottom two panels of Figure~\ref{fig:scaling_factors}]{Miville-Desch2005}. There is no trend between \rxs\ and galactic latitude. There is also no trend of \rxs{} and IRAS surface brightness; however, two of the clusters -- Abell~478 and PKS~0745$-$191 -- are notable outliers in their IRAS surface brightness. Removing these two clusters from our cosmological analysis has a negligible impact on the results (Section~\ref{sec:H0}).

We performed an additional test to determine if the results depend on the exact algorithm used to identify and remove potential contamination. For the three low-redshift clusters where pixels were removed, we re-ran the analysis using S/N thresholds of 4 and 6 (compared to our baseline of 5). The only cluster with a non-negligible change in the best-fitting \rxs\ was Abell 478, which varied by approximately 4 per cent based on the S/N choice. We verified that this difference is due to an extended feature with multiple pixels near the S/N threshold. In order to account for this, we increased the measurement error for this cluster by including additional Gaussian noise with a standard deviation of 4 per cent. 

\paragraph{Projected clusters:}
The probability of another massive cluster falling within a projected radius of 6\,$r_{500}$ of a given target cluster is small but non-zero, particularly for the lowest-redshift objects in our sample. As noted in Section~\ref{sec:SZ_data}, we searched for nearby clusters using the MCXC catalog of \citet{Piffaretti2011}, finding two such examples. Regions centered on these two objects were removed from the $y$ maps used in our analysis. The value of \rxs\ decreased by $\sim 9$ per cent in Abell~2029 and by $\sim 0.7$ per cent in MS~2137.3$-$2353 following the removal of these regions. Furthermore, any potential clusters not in the MCXC with bright SZ effect signals will be excluded from our analysis based on our removal of pixels away from the clusters' centers with a S/N~$\ge 5$.

\subsubsection{Calibration uncertainties}
\label{sec:calibration}

\paragraph{X-ray measurements:}
\label{sec:xraycal}
We consider the {\it Chandra} calibration to consist of two distinct components: the accuracy of the instrument response model at soft energies ($\ltsim2$\,keV, the primary determinant of gas density measurements for clusters in the relevant temperature range for this work) and the accuracy of temperature determinations from spectral fits of the bremsstrahlung continuum.
Compared with the response as a function of energy, this simplification is much more amenable to being characterized using on-orbit observations of celestial sources.
In particular, the soft response of {\it Chandra} and XMM-{\it Newton} are known to agree at the per cent level \citep{Nevalainen2010, Schellenberger2015}; correspondingly, gas density determinations from the two telescopes are in good agreement \citep{Rozo2014}.
Taking this as the scale of the overall flux calibration uncertainty results in an estimated bias in \rxs{} at the $\ltsim1$ per cent level, which is negligible compared with other systematics.

In contrast, temperatures measured by {\it Chandra} and XMM-{\it Newton} are known to be discrepant, particularly for hot clusters like those used in this work \citep{Nevalainen2010, Schellenberger2015}.
While the existence of a discrepancy does not inform us about the correct absolute calibration, we can construct a prior for the temperature calibration by combining X-ray observations with information from simulations and weak gravitational lensing measurements, as follows.
Cluster total masses estimated from X-ray measurements of gas density and temperature, assuming hydrostatic equilibrium (HSE), may be biased for two principal reasons: violations of HSE and bias in the adopted temperature calibration (with mass directly proportional to temperature in the hydrostatic equation). 
The former is minimized by measuring masses of the most dynamically relaxed clusters at intermediate radii (though it will still be present at some level), while the second is precisely what we aim to quantify.
With careful control of systematics, weak lensing masses are expected to be unbiased at the per cent level (e.g.,\ \citealt{Applegate2014}).
Thus, an expectation for the mass bias due to departures from HSE in relaxed clusters from hydrodynamic simulations can be combined with comparisons of biased X-ray and unbiased weak lensing mass estimates to constrain the X-ray temperature calibration.
To be explicit, we have
\begin{equation}
    1+\bkt = \frac{1}{\rwlx (1+\bhse)},
    \label{eq:temp bias}
\end{equation}
where $\rwlx$ is the ratio of weak lensing to X-ray mass, $\bhse$ is the X-ray hydrostatic mass bias, and $\bkt$ is the temperature bias.

We adopt a uniform prior between $-0.1$ to 0.0 for $\bhse$, a range comfortably including the predictions of hydrodynamical simulations specifically of massive, relaxed clusters at radii $\sim r_{2500}$ \citep{Nagai2007, Ansarifard2020}.
The ratio of weak lensing to X-ray masses at the same radius for clusters satisfying the same relaxation criterion as our sample was measured to be $\rwlx=0.96\pm0.12$, accounting for both statistical and lensing systematic uncertainties \citep{Applegate2016}.\footnote{\citet{Applegate2016} employed X-ray temperatures and masses from \citet{Mantz1509.01322}, which are nearly identical to those in this work (Section~\ref{sec:xdata}).}
Straightforwardly combining this information leads to an estimated X-ray temperature bias of $\bkt=0.09 \pm 0.13$, where the positive sense implies an overestimate.
Our cosmological analysis marginalizes over this uncertainty, which dominates the final error budget for the $H_0$ measurement.

We emphasize that the nominal value of 0.09 for hot ($kT\gtsim5$\,keV) clusters is specific to {\it Chandra} (and {\sc caldb} versions that produce temperatures consistent with ours), but the uncertainty is not.
Indeed, the width of 13 per cent is likely to be the highest precision possible at the present time, given that our restriction to massive, relaxed clusters minimizes both the expected hydrostatic bias and its uncertainty from simulations, while the weak lensing to X-ray mass ratio constraint was obtained for exactly this class of clusters.
There are, however, several routes to improving the precision in the future, which we discuss in Section~\ref{sec:future}.

An additional subtlety is that our X-ray temperature estimates were used to account for the relativistic corrections to the SZ spectral distortion while jointly fitting the pressure profiles (see Section~\ref{sec:deprojection}).
We found that the net impact on \rxs{} of adjusting the gas temperature by an overall multiplicative factor is well described by a linear function.
The precise slope varies from cluster to cluster with the shape of the temperature and pressure profiles and the radial dependence of the X-ray and SZ pressure measurement uncertainties.
This dependence, which slightly reduces the impact of the temperature calibration on the inferred value of $H_0$, was also accounted for in our cosmological fit (see Section~\ref{sec:model}).

\paragraph{SZ measurements:}
As detailed in Section~\ref{sec:SZ_data}, the {\it Planck} absolute calibration uncertainty is 0.1 per cent. The Bolocam data are referenced to the {\it Planck}-calibrated planetary model, which has an absolute calibration uncertainty of 0.6 per cent. These uncertainties are negligible compared with other systematics discussed above. The random 1.6 per cent variation in the Bolocam calibration relative to the absolute standard is included in the noise realizations during our pressure profile fits. 

\paragraph{Fitting procedure:}
\label{sec:fit_bias}
As discussed in Section~\ref{sec:deprojection}, approximations made in our pressure profile fitting procedure result in an overall bias of $0.024\pm0.026$ in $\ln\rxs{}$.

\subsection{Complete Likelihood}
\label{sec:model}

In addition to cosmological parameters, our model contains 3 nuisance parameters: a log-normal intrinsic scatter, \sint, accounting for cluster-to-cluster variations in \rxs{}; an overall bias in X-ray temperature measurements, \bkt; and an overall bias (excluding that due to $\bkt$) impacting measurements of $\ln\rxs$, \bF.
The latter two are parametrized separately because \bkt{} impacts both the X-ray and SZ signals in such a way that a given bias in temperature does not translate directly to the same bias in \rxs{} (see Section~\ref{sec:xraycal}).
Our baseline analysis assumes a flat $\Lambda$CDM cosmology with $\Omegam=0.3$, $\Omegal=0.7$ and $H_0$ as a free parameter (our sensitivity to $\Omegam$ is small enough that the precise value assumed is unimportant).
We employ a uniform prior on $H_0$ between 0 and 200\,\kmsMpc, a minimally informative Jeffreys prior on the intrinsic scatter ($\propto \sint^{-1}$), and Gaussian priors $\bkt=0.09\pm0.13$ (Section~\ref{sec:xraycal}) and $\bF=0.024\pm0.026$ (Section~\ref{sec:fit_bias}).
Extensions to this baseline model, using different priors, will be investigated in Section~\ref{sec:results}.

\begin{figure}
	\includegraphics[width=\columnwidth]{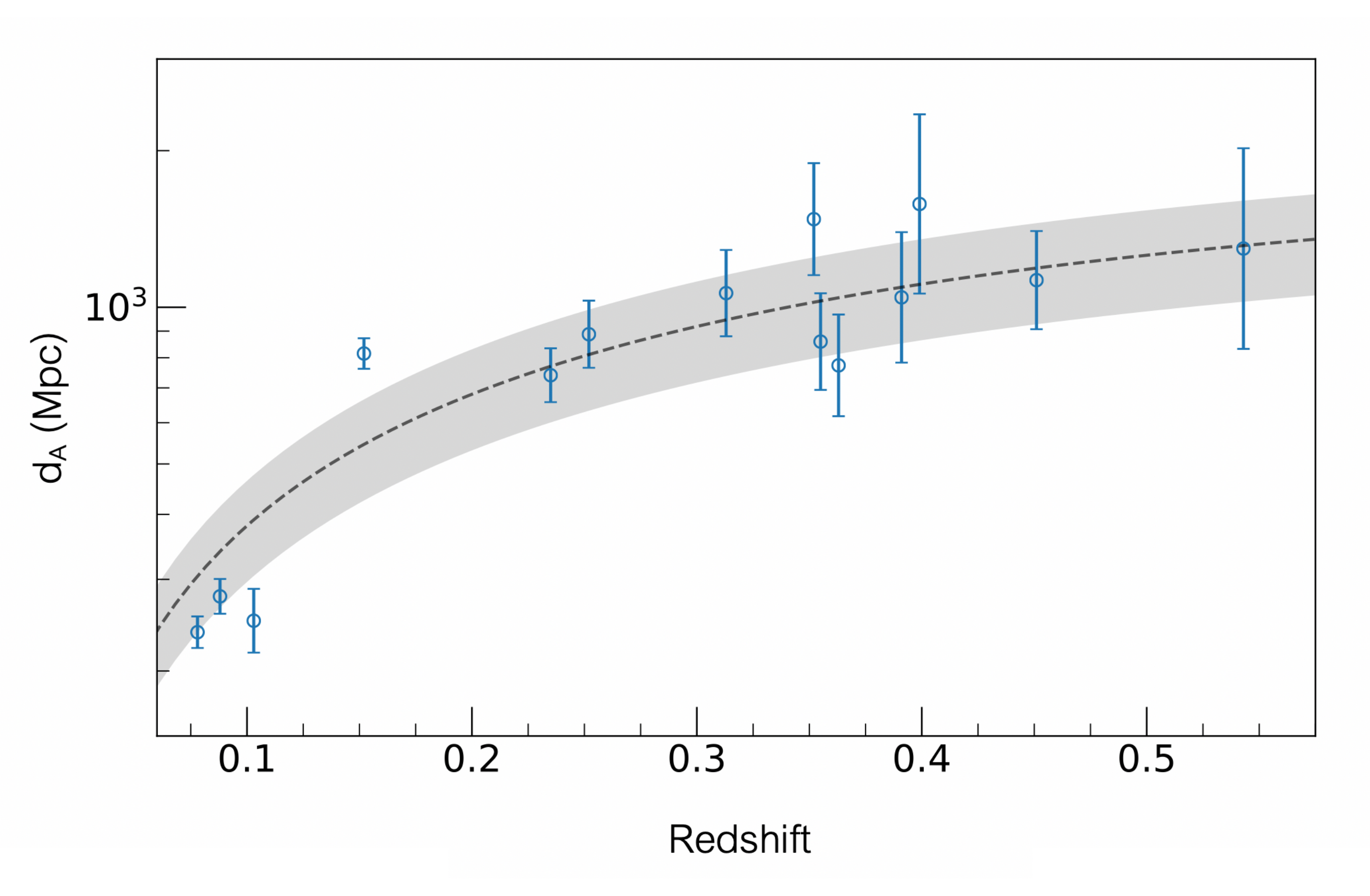}
	\caption{
	Angular diameter distance estimated from each cluster as a function of redshift.
	The dashed, black line represents the relation between $\dA$ and $z$ predicted by a concordance model ($H_0=70\,\kmsMpc$, $\Omegam = 0.3$, $\Omega_{\Lambda} = 0.7$), with the shaded gray region indicating our best-fitting intrinsic scatter. 
    }
    \label{fig:DA_vs_z}
\end{figure}

The likelihood function can be written as
\begin{equation}
    \mathcal{L}(H_0, \sint, \bkt, \bF) = \prod_i \int ds \, p_i(\ln\rxs_i | s) p(s | \smean_i, \sint),
	\label{eq:likelihood}
\end{equation}
where the product is over clusters in the data set, $p_i(\ln\rxs_i | s)$ is the non-central $t$ distribution describing the measurement uncertainty for the $i$th cluster (Section~\ref{sec:deprojection}), and $p(s | \smean_i, \sint)$ is the log-normal density for the given mean and standard deviation.
The expectation value, $\smean_i$, is given by
\begin{equation}
    \smean_i = \frac{1}{2}\ln\left[\frac{\dref(z_i)}{\dA(z_i)} \right] + \ln(1 + \alpha_i \bkt)  + \bF,
	\label{eq:s_mean}
\end{equation}
where $d(z)$ is the angular diameter distance as a function of redshift, $\dref(z)$ is the distance as estimated in the reference cosmology assumed when measuring \rxs{} from the X-ray and SZ data (Section~\ref{sec:cosmodel}), and $\alpha_i$ is the linear slope relating the estimate of \rxs{} with the temperature calibration bias for the $i$th cluster.
Figure~\ref{fig:DA_vs_z} shows the angular diameter distances estimated for each cluster by inverting Equation~\ref{eq:s_mean} and propagating the statistical uncertainties in \rxs{}, for nominal values of \bkt{} and \bF{} (below), and compares them with predictions from a concordance $\Lambda$CDM model.

Our constraints were obtained using the {\sc emcee} Markov chain Monte Carlo code \citep{Foreman-Mackey2013}.

\section{Results} \label{sec:results}

In this section, we report and discuss our constraints on cosmological parameters and intrinsic scatter using the baseline priors defined in Section~\ref{sec:model} (summarized in Table~\ref{tab:results}), and, conversely, implications for the X-ray temperature bias obtained from external priors on $H_0$. 

\begin{table*}
	\centering
	\caption{
	Marginalized 68.3 per cent confidence intervals on $H_0$ and the intrinsic scatter parameter of our model.
	We employ a uniform prior on $H_0$ between 0 and 200 \kmsMpc, a minimally informative Jeffreys prior on the intrinsic scatter ($\propto \sint^{-1}$), and Gaussian priors $\bkt = 0.09 \pm 13$ (except where noted) and $\bF = 0.024 \pm 0.026$.
	$\Omegam$ is fixed to 0.3 by default, and marginalized over a uniform prior from 0 to 1 when free.
	}
	\label{tab:results}
	\begin{tabular}{lcc}
	    \hline
		\hline
		& $H_0$ (\kmsMpc) & \sint \\
		\hline
		Baseline analysis & $67.3^{+21.3}_{-13.3}$ & $0.13 \pm 0.04$\vspace{1ex}\\
		Excluding A478 \& PKS 0745 & $65.7^{+16.8}_{-13.3}$ & $0.10 \pm 0.03$\vspace{1ex}\\
		With \Omegam\ free & $66.5^{+21.1}_{-13.4}$ & $0.13 \pm 0.04$\vspace{1ex}\\
		Fixing $\bkt = 0.09$ & $72.3^{+7.6}_{-7.6}$ & $0.12 \pm 0.04$\vspace{1ex}\\
		Fixing $\bkt = 0.0$ & $84.7^{+8.9}_{-9.1}$ & $0.13 \pm 0.04$\\
		\hline
	\end{tabular}
\end{table*}

\begin{figure}
	\includegraphics[width=\columnwidth]{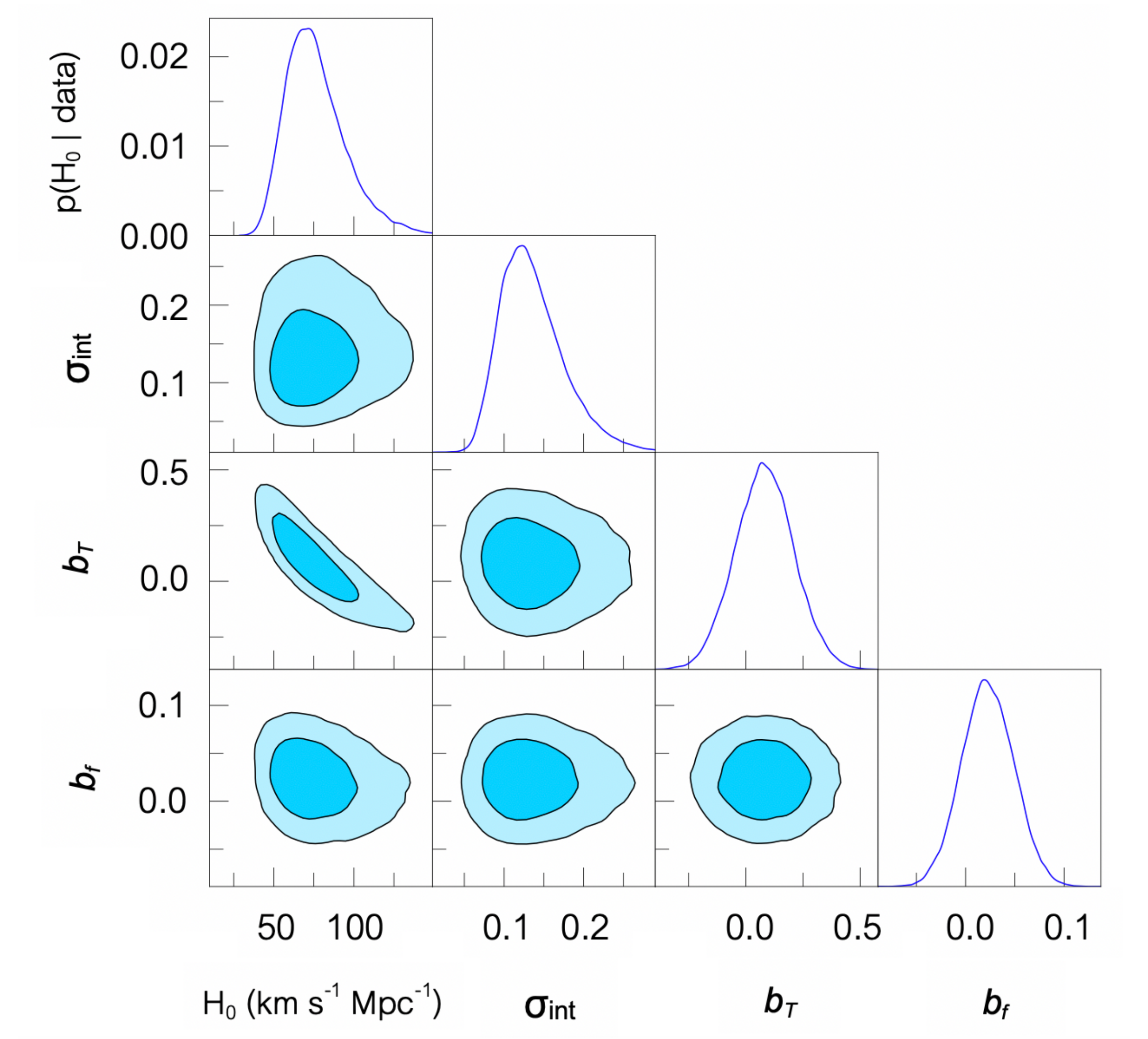}
	\caption{Parameter constraints from our baseline analysis.
	Panels on the diagonal show 1-dimensional, marginalized posteriors for each parameter, while off-diagonal panels show joint, marginalized 68.3 and 95.4 per cent credible regions for each parameter pair.
    }
    \label{fig:triangle}
\end{figure}

\subsection{Hubble parameter} \label{sec:H0}

Figure~\ref{fig:triangle} shows constraints on the model parameters from our baseline analysis, which yields $H_0 = 67.3^{+21.3}_{-13.3}\,\kmsMpc$.
As the figure indicates, there is a strong degeneracy between $H_0$ and \bkt{}, with uncertainty in the X-ray temperature calibration limiting the precision of the $H_0$ determination from this method.
Our results do not change appreciably when excluding Abell~478 and PKS~0745$-$191, the clusters with the largest potential dust contamination, from the analysis ($65.7^{+16.8}_{-13.3}$\,\kmsMpc), nor when marginalizing over flat \LCDM{} models with $\Omegam$ as a free parameter ($66.5^{+21.1}_{-13.4}$\,\kmsMpc; in this case we also obtain the upper limit $\Omegam<0.55$ at 68.3 per cent confidence).

\begin{figure*}
\centering
\begin{minipage}{0.5\textwidth}
  \centering
  \includegraphics[width=\textwidth]{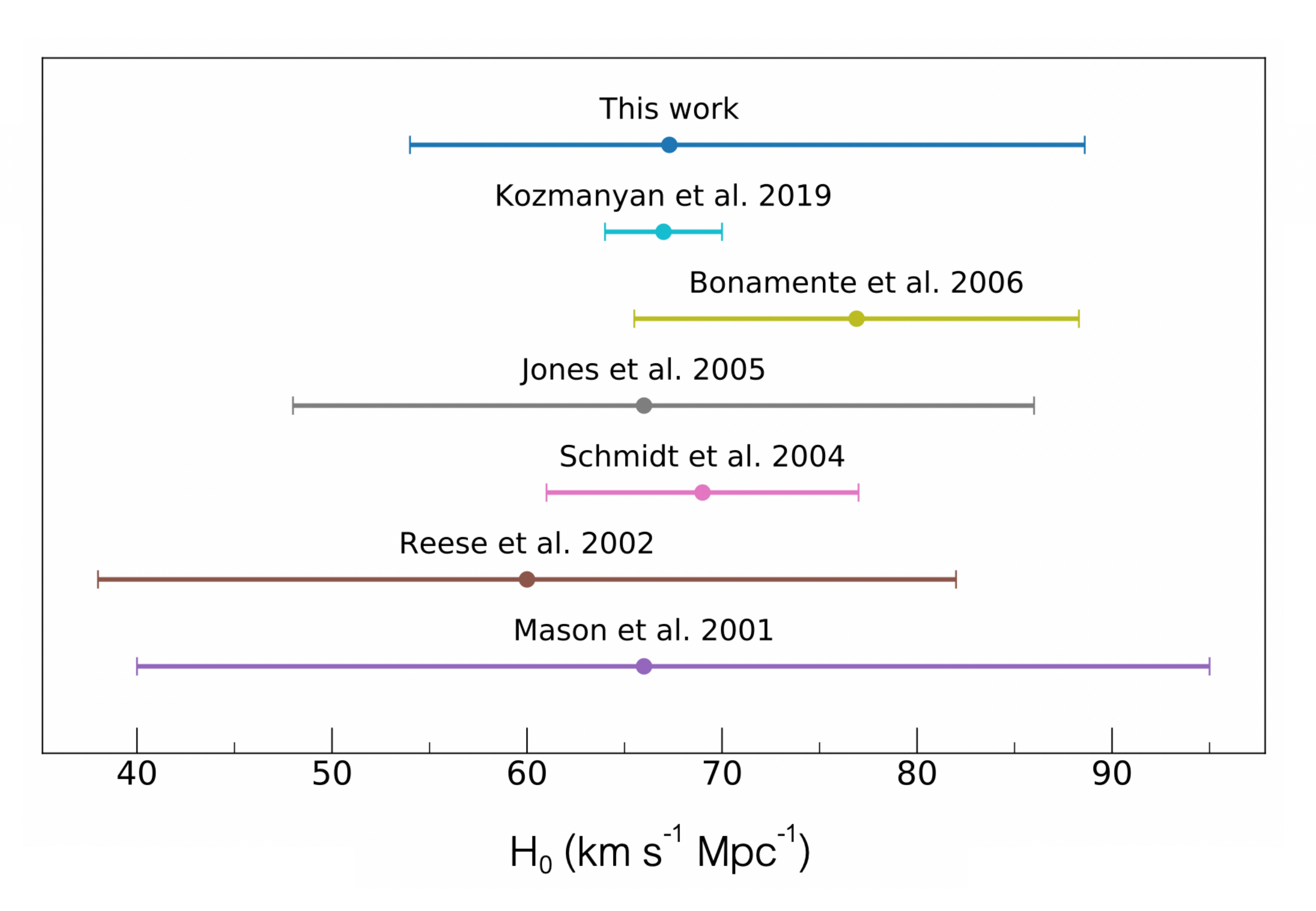}
\end{minipage}%
\begin{minipage}{0.5\textwidth}
  \centering
  \includegraphics[width=\textwidth]{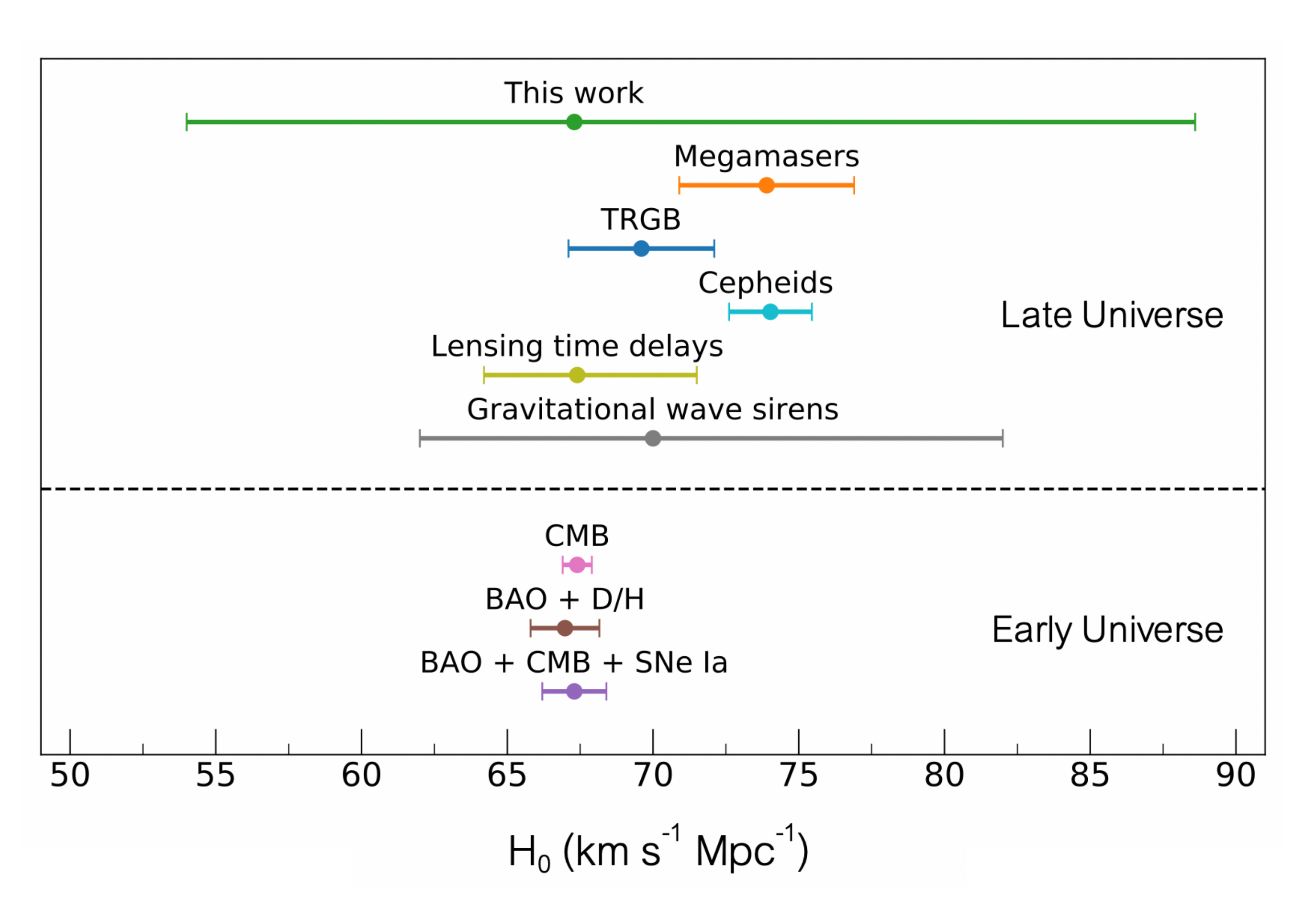}
\end{minipage}
\caption{Comparison of the result in this work with previous measurements of $H_0$ using clusters as a probe (left) and other cosmological probes (right).
Error bars reflect the statistical+systematic uncertainty as reported in each work.
In the right panel, the $H_0$ values are split by early Universe (CMB, \protect\citealt{Planck2018_VI}; BAO+D/H, \protect\citealt{Addison2018}; BAO+CMB+SNeIa, \protect\citealt{Aubourg2015}) and late Universe (this work; GW sirens, \protect\citealt{Abbott2017}; lensing, \protect\citealt{Birrer2020}; Cepheids, \protect\citealt{Riess2019}; TRGB, \protect\citealt{Freedman2020}; megamasers, \protect\citealt{Pesce2020}) measurements.
}
\label{fig:H0_comparison}
\end{figure*}

In Figure~\ref{fig:H0_comparison} (left panel), we show our $H_0$ measurement alongside a selection of previous measurements made using X-ray and SZ observations of clusters.
The error bars for these works reflect the total statistical+systematic uncertainty reported by the authors, but we note that estimates of the applicable systematic uncertainties, and even what effects are included, vary greatly.
In particular, the relatively precise results of \citet{Kozmanyan2019} do not account for any systematic uncertainty in X-ray temperature measurements, while earlier works typically use optimistic estimates based on discrepancies between {\it Chandra} and XMM-{\it Newton}, or between different versions of their respective calibration files; the present study is the first to employ empirically determined limits on this key systematic effect.
It is also worth noting that the majority of previous works were not restricted to the most dynamically relaxed systems, increasing systematic uncertainties related to dynamical activity such as mergers and departures from ellipsoidal symmetry.

The right panel of Figure~\ref{fig:H0_comparison} compares our results with those of independent cosmological probes, which we conventionally organize into groups sensitive to ``early Universe'' and ``late Universe'' physics.
Amongst the former are analyses of CMB primary anisotropy and lensing data, for which precise constraints on $H_0$ are possible only when assuming a flat $\Lambda$CDM model \citep{Planck2018_VI};
and BAO combined with external information on either the cosmic baryon density (from primordial deuterium abundances, BAO+D/H; \citealt{Addison2018}), or the acoustic scale at recombination and the cosmic expansion history (BAO+CMB+SNeIa; \citealt{Aubourg2015}). The late-Universe probes shown include traditional $H_0$ estimates from the distance ladder, calibrated using Cepheid variables \citep{Riess2019} or the tip of the red giant branch \citep[TRGB;][]{Freedman2020};
gravitational lensing time delays \citep{Birrer2020}; gravitational wave sirens \citep{Abbott2017}; and distance measurements to megamaser-hosting galaxies \citep{Pesce2020}. 

\begin{figure}
	\includegraphics[width=\columnwidth]{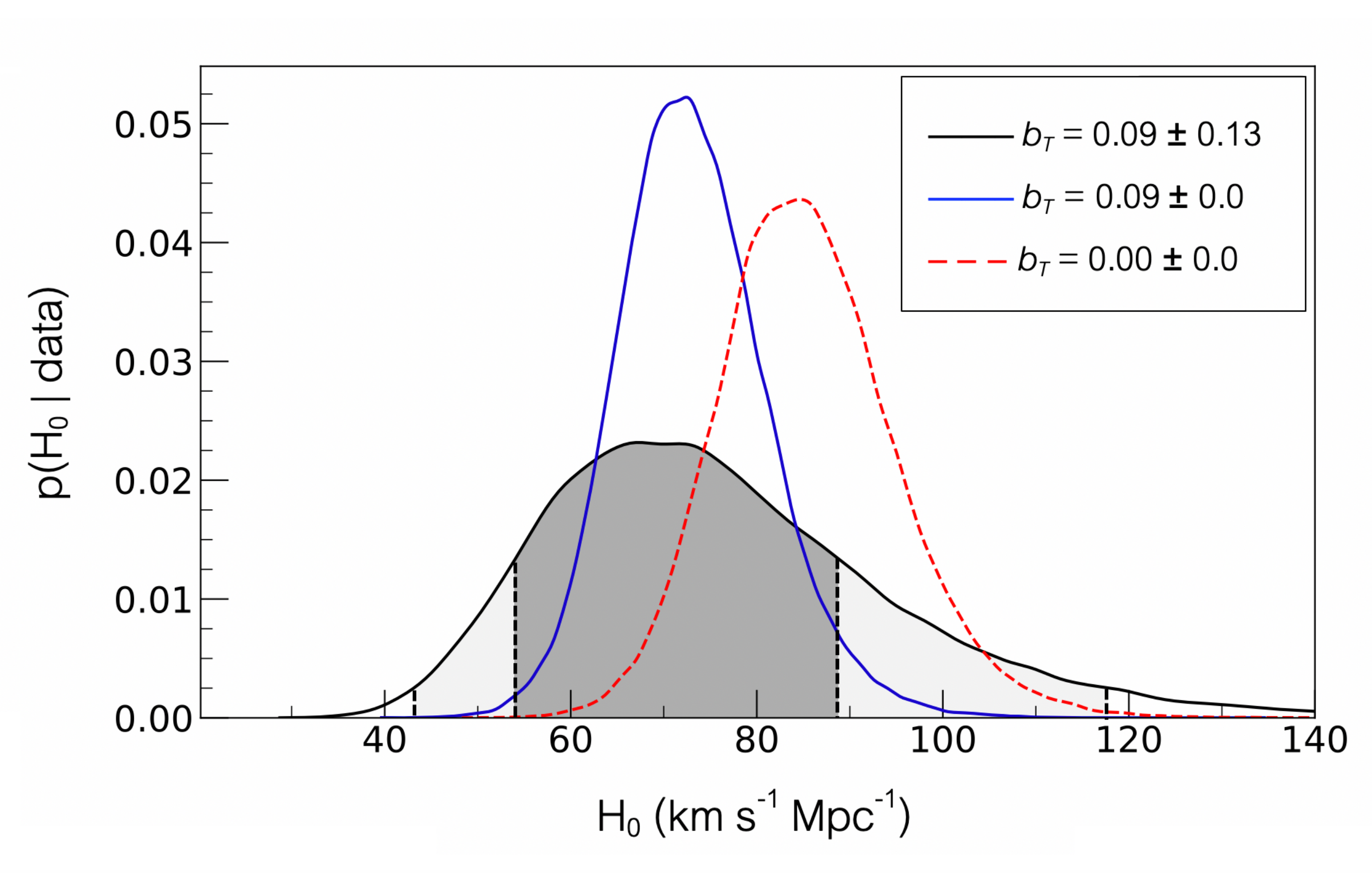}
	\caption{Posterior distributions of $H_0$ for a flat $\Lambda$CDM universe with varying values of X-ray temperature calibration bias \bkt. Dark and light shaded regions show the marginalized 68.3 per cent and 95.4 per cent confidence intervals when marginalizing over a Gaussian prior $\bkt=0.09\pm0.13$. In the other cases shown, \bkt{} is fixed at the indicated values.
	}
    \label{fig:H0_pdfs}
\end{figure}

The dominant limitation of our constraint on $H_0$ is the systematic uncertainty in X-ray temperature measurements.
Section~\ref{sec:future} discusses the possibilities for reducing this uncertainty, and the corresponding prospects for improvement in cosmological constraints.
For now, we note, as points of reference, the constraints from our analysis when the temperature calibration parameter is held fixed at either $\bkt=0.09$ (the center of our prior) or $\bkt=0.0$ (perfectly calibrated data).
Respectively, these assumptions yield $H_0 = 72.3^{+7.6}_{-7.6}\,\kmsMpc$ and $H_0 = 84.7^{+8.9}_{-9.1}\,\kmsMpc$ (Figure~\ref{fig:H0_pdfs}), where the uncertainties are now primarily statistical (limited by the sample size given the intrinsic cluster-to-cluster scatter) rather than systematic.
Looking at this another way, the precise measurements of $H_0$ available from some other techniques, in order to be consistent with the SZ data, imply a calibration bias resulting in overestimates of the X-ray temperature at the $\sim5$--15 per cent level.
We explore the implications for the temperature calibration based on using this kind of approach in more detail in Section~\ref{sec:bias_constraints}.

\subsection{Intrinsic scatter}

Our baseline analysis constrains the intrinsic scatter in $\ln\rxs{}$ to be $\sint = 0.13 \pm 0.04$, corresponding to a $26\pm8$ per cent scatter in the value of $H_0$ inferred from a single cluster (Equation~\ref{eq:s_mean}).
This parameter subsumes any systematic scatter due to effects listed in Section~\ref{sec:systematics} that were not incorporated into the noise realizations used to quantify our measurement errors, but we expect the dominant contributor to $\sint$ to be variation in the viewing angle of non-spherical clusters.
Semi-analytic estimates of the scatter due to asphericity by \citet{Sulkanen1999} and \citet{Kozmanyan2019} are broadly consistent with our findings.
As noted previously, additional scatter due to, e.g., gas clumping \citep{Kozmanyan2019}, is likely to be minimized due to our sample selection and the primary sensitivity of our measurements to intermediate radii where such astrophysical effects are small.
When Abell~478 and PKS~0745$-$191, the clusters with the largest expected dust contamination, are excluded, we find a scatter of $\sint  = 0.10 \pm 0.03$.

\subsection{X-ray temperature and HSE biases} \label{sec:bias_constraints}

\begin{figure}
	\includegraphics[width=\columnwidth]{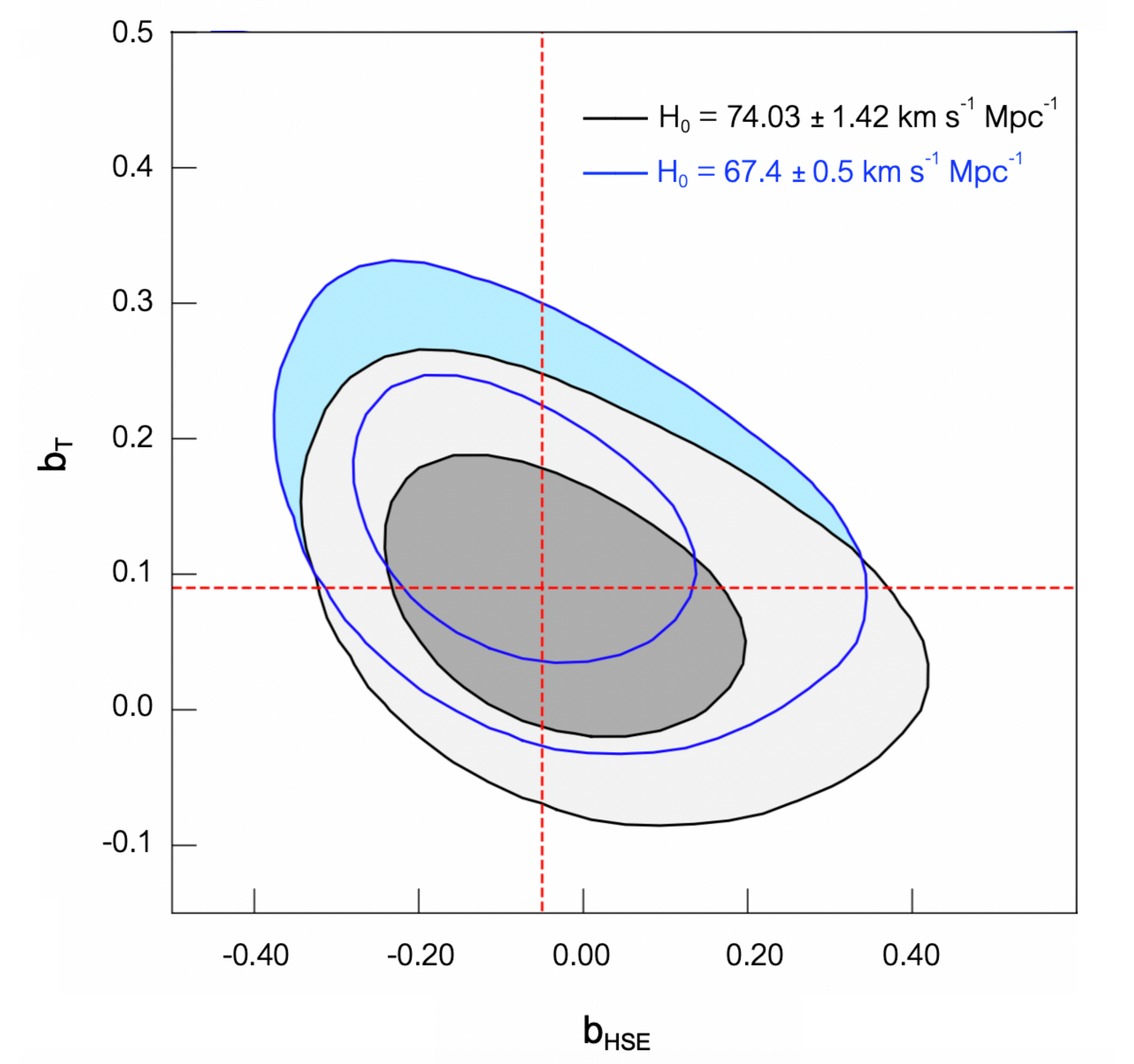}
	\caption{Constraints on the X-ray temperature bias, \bkt, and the implied HSE mass bias, $b_\mathrm{HSE}$, using $H_0$ priors derived from \protect\cite{Riess2019} and the \protect\cite{Planck2018_VI}. The dark and light shaded regions show the marginalized 68.3 per cent and 95.4 per cent confidence regions. Red dashed lines indicate the central values of the priors used for each parameter in the baseline analysis.
    }
    \label{fig:kT_vs_HSE_bias}
\end{figure}

In this section, we explore the implications for the X-ray temperature calibration and HSE mass bias if we adopt priors on $H_0$ from external data.
We specifically consider the constraints reported by \citet[][ $H_0=(74.03\pm1.42)\,\kmsMpc$]{Riess2019} and the \citet[][$H_0=(67.4\pm0.5)\,\kmsMpc$]{Planck2018_VI}.
Figure~\ref{fig:kT_vs_HSE_bias} shows the joint constraints on the X-ray temperature calibration parameter and the hydrostatic mass bias, the two being related via the measured X-ray mass to weak lensing mass ratio from \citet[][see Equation~\ref{eq:temp bias} in Section~\ref{sec:xraycal}]{Applegate2016}.
Using the $H_0$ prior based on \citet{Riess2019}, the individual, marginalized constraints are $\bkt = 0.08 \pm 0.06$ and $\bhse = -0.04 \pm 0.13$.
Employing the Planck prior instead, we find $\bkt = 0.13 \pm 0.07$ and $\bhse = -0.09 \pm 0.13$.
As mentioned in Section~\ref{sec:H0}, values of $H_0$ consistent with external data favor an overestimated temperature calibration at modest confidence; the implied values of the HSE mass bias are in good agreement with predictions from hydrodynamical simulations (Section~\ref{sec:xraycal}). 
We note that these results are specific to {\it Chandra} and the version of the calibration files used in this analysis.
The modest degeneracy of the 2-dimensional constraints in Figure~\ref{fig:kT_vs_HSE_bias} indicates that the precision of our results on \bhse{} is limited by the precision on the weak lensing to X-ray hydrostatic mass ratio,
rather than the precision of the $\bkt$ constraint (which itself follows from $H_0$ for the analysis described in this section).

\section{Future prospects} \label{sec:future}

While constraints on $H_0$ from the method employed here are systematically limited to lower precision than many published constraints from other probes, there is reason to be optimistic that the primary systematic uncertainty, due to X-ray temperature calibration, can be greatly reduced.
Presently the temperature calibration is constrained to 13 per cent; recall that the resultant fractional uncertainty on $H_0$ is approximately doubled (when other cosmological parameters are held fixed; Equation~\ref{eq:s_mean}) to 26 per cent.

In the near term, constraints on the temperature calibration, using our method of combining a simulation prior on the X-ray hydrostatic mass bias with weak lensing mass measurements, can be straightforwardly tightened by a factor of $\sim2$ by expanding the amount of high-quality lensing data (with sufficient color coverage to obtain robust photometric redshifts) for the relaxed cluster sample.
A more ambitious, but entirely feasible, approach is to enable more direct in-situ calibration by launching a well calibrated X-ray source into orbit.
For example, the Cal X-1 mission concept, consisting of a source-telescope pair of CubeSats, could provide the basis for robust on-orbit calibration of X-ray detectors as a function of energy, yielding an effective temperature calibration at the $\sim3$ per cent level \citep{Jahoda2019}.

Another option is to calibrate temperatures measured from X-ray observations to temperature measurements based on the relativistic corrections to the thermal SZ effect sourced by the same gas. The challenge of this approach is that the relativistic corrections are generally small, and are most easily observed at higher frequencies ($\gtrsim 300$\,GHz). At these frequencies, Doppler-broadened line emission from water vapor in the atmosphere presents a challenge to ground-based observations. In addition, the thermal dust emission from individual background galaxies typically dwarfs the SZ effect signal, and sufficient angular resolution and spectral coverage are required to subtract this emission \citep[see, e.g.,][]{Sayers2019}.

High quality space-based data that address these challenges are available from {\it Herschel}-SPIRE for a sample of approximately 50 clusters \citep{Egami2010,Oliver2012}, and work is currently underway to measure ICM temperatures with the SZ effect using these data. We estimate that the resulting temperature calibration will have a statistical uncertainty of $\simeq 3$ per cent over the range $\sim 8$--12\,keV. Outside of this range, statistical uncertainties of $\lesssim 5$ per  cent should be achievable down to temperatures as low as $\sim 6$\,keV. Such a measurement would also include systematic uncertainties related to the SPIRE calibration \citep[$\simeq 4$ per cent, see][]{Bertincourt2016}, and potential systematics due to non-uniform and non-isothermal ICM distributions \citep[$\simeq 1$--5 per cent based on simulations, e.g.][]{Biffi2014,Morandi2013}. In sum, this calibration scheme could potentially result in a temperature calibration accurate to 5 per cent.

In the longer term, future large-aperture, multi-band observations have the potential to provide deprojected density profiles from SZ data alone, by simultaneously constraining both pressure and temperature using the relativistic corrections to the thermal SZ effect \citep{Morandi2013,Klaassen2020}. Such data would obviate the measurement of X-ray temperatures in the first place: given the insensitivity of X-ray densities to temperature (for the temperature range of interest), Hubble parameter constraints could then be obtained by comparing estimates of ICM density rather than pressure.
As noted in Section~\ref{sec:calibration}, X-ray and SZ effect data have already achieved (sub) per cent level flux calibration, and so percent-level $H_0$ constraints may be possible with such data. However, we note that, while the South Pole Telescope and the Atacama Cosmology Telescope have demonstrated ground-based absolute flux calibration uncertainties of $\lesssim 0.5$ per cent at 90 and 150~GHz \citep{Hou2018,Choi2020}, it is more difficult to obtain such precision at the higher frequencies necessary to measure the relativistic corrections to the SZ effect \citep[e.g.,][]{Louis2014,Sayers2019}.

\begin{figure}
	\includegraphics[width=\columnwidth]{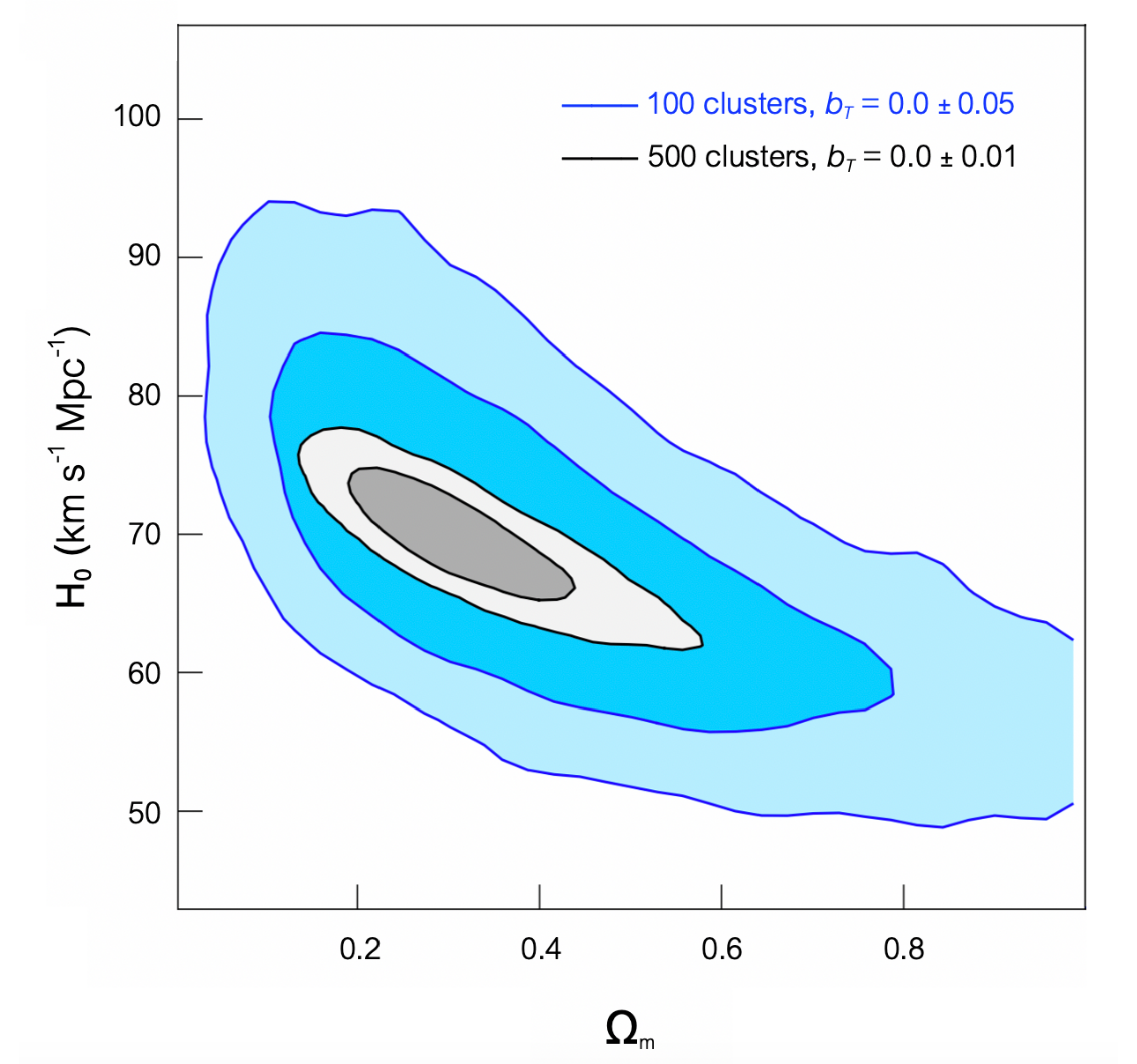}
	\caption{Forecasted constraints on $\Omegam$ and $H_0$ from mock cluster data (see Section~\ref{sec:future}).
	}
    \label{fig:OmH0}
\end{figure}

While our current constraints on $H_0$ are essentially unchanged when more freedom is allowed in the cosmological model (in the form of leaving $\Omegam$ free), this will not be the case for future measurements with reduced temperature calibration uncertainty.
Figure~\ref{fig:OmH0} shows joint constraints on $H_0$ and $\Omegam$ (for flat $\Lambda$CDM models) forecasted for two benchmark analyses, following the procedure of \cite{Allen2013}.
In both scenarios, we assume an intrinsic scatter based on our best estimate, $\sint=0.13$, along with measurement uncertainties subordinate to the intrinsic scatter.
Clusters in each mock data set were distributed in redshift according to the halo mass function, with an additional restriction to average temperatures $kT>5$\,keV (as for the present data set).
In the first case (blue contours), we assume a sample of 100 clusters, with a temperature calibration accurate to 5 per cent, corresponding to the near-term expectation.
Gray contours show constraints from 500 clusters with a longer-term temperature calibration accurate to 1 per cent.
We find improved constraints on $\Omegam$ from these forecasts compared with those of \citet{Allen2013}, due to their assumption of a larger intrinsic scatter ($\sint=0.2$) than we find.
However, we note that a sample as large as 500 may require loosening the selection criteria, and the extent to which that can be done without increasing the intrinsic scatter will need to be studied.

The forecasted constraints from both scenarios display a degeneracy between $H_0$ and $\Omegam$, reflecting the lack of clusters at extremely low redshifts where $d(z)$ is approximately linear.
In practice, the cosmological parameters describing the shape of $d(z)$, such as $\Omegam$, are efficiently constrained using other methods.
In particular, we note the synergy with cosmological tests based on the cluster gas mass fraction \citep{Mantz1402.6212}, which employ exactly the same X-ray data set (imaging spectroscopy of hot, dynamically relaxed clusters).
These data are sensitive to relative distances as $d(z)^{3/2}$, as well as the combination $H_0^{3/2} \Omega_\mathrm{b} / \Omegam$ through the absolute value of the gas mass fraction, the latter providing a route to tight constraints on $\Omegam$ even in general dark energy models.
Along with independent relative distance proxies \citep{Kim1309.5382}, we can expect such data to break degeneracies like that shown in Figure~\ref{fig:OmH0}, meaning that the simple relation $\Delta H_0 \simeq 2\Delta \bkt$ provides a good estimate of potential future constraints on $H_0$ from X-ray and SZ cluster data.
The various possibilities discussed above thus correspond to precisions of $\sim12$ 
per cent on $H_0$ in the near term (from improved weak lensing data) to $\sim1$ per cent in the longer term (from resolved relativistic SZ measurements).

\section{Conclusion} \label{sec:conclusion}

We obtained constraints on the Hubble parameter from the combination of {\it Chandra} X-ray and Bolocam and/or {\it Planck} SZ effect observations of 14 massive, dynamically relaxed clusters.
Our constraints arise from the relative normalization of the ICM pressure profiles inferred from X-ray and SZ data, along with X-ray pressure estimates depending on distance as $d(z)^{-1/2}$. We fit these data at intermediate radii where astrophysical uncertainties due to feedback (at small radii) and clumping/accretion (at large radii) are minimized.
Our model includes the intrinsic scatter in this X-ray/SZ pressure ratio as a free parameter, and marginalizes over systematic uncertainties due to calibration and other sources.
The dominant uncertainty impacting our constraint on $H_0$ is that of the accuracy of ICM temperatures determined from X-ray data, for which we derive a prior by combining weak lensing and X-ray hydrostatic mass estimates with the simulation expectation for the mass bias in the latter.
This marks the first time that the primary systematic uncertainty impacting this cosmological test has been fully and empirically accounted for. 
Our findings can be summarized as follows.
\begin{enumerate}
    
    \item Assuming a flat \LCDM{} model with $\Omegam=0.3$ and $\Omegal=0.7$, we find $H_0=67.3^{+21.3}_{-13.3}$\,\kmsMpc.
    The quoted errors are dominated by the systematic uncertainty in X-ray temperature estimates; neglecting this effect (fixing the associated nuisance parameter to its nominal value), we would obtain a statistically limited constraint of $H_0 = 72.3^{+7.6}_{-7.6}$\,\kmsMpc.
    Our results are in good agreement with previous estimates using this technique, and with independent cosmological probes.
    
    \item The log-normal scatter of the X-ray/SZ pressure ratio is found to be $\sint=0.13\pm0.04$, corresponding to a $26\pm8$ per cent scatter in the distance estimated from a single cluster.
    This is broadly consistent with the expected impact of cluster asphericity, which is likely to be the primary source of scatter for the data set employed here (relaxed clusters, resolved in both the X-ray and SZ data and constrained at intermediate radii).
    Excluding the two clusters whose SZ data are most likely to suffer from residual dust contamination, the scatter is $\sint=0.10\pm0.03$.
    
    \item When adopting an external prior on $H_0$, our data can be used along with measurements of the weak lensing to X-ray hydrostatic mass ratio to infer both the overall calibration of the X-ray temperature measurements and the intrinsic bias in X-ray masses due to departures from equilibrium (for relaxed clusters).
    Using an $H_0$ prior from the CMB, we find a temperature calibration parameter of $\bkt = 0.13 \pm 0.07$.
    With an $H_0$ prior based on the Cepheid-calibrated distance ladder, we find $\bkt=0.08\pm0.06$.
    Put differently, for values of $H_0$ in the commonly accepted range, the SZ data imply that the X-ray temperatures of hot ($kT\gtsim5$\,keV) clusters are overestimated at the $\sim10$ per cent level (specifically for {\it Chandra} at the {\sc caldb} version we used), albeit only at 1--$2\sigma$ confidence.
    The corresponding estimates for the hydrostatic mass bias are, respectively, $-9\pm13$ and $-4\pm13$ per cent, in good agreement with simulations.
    
    \item Several approaches can potentially reduce the systematic uncertainties currently limiting constraints on $H_0$ using X-ray and SZ cluster data, including more precise measurements of the weak lensing to X-ray hydrostatic mass ratio for relaxed clusters, calibration of X-ray temperature measurements using known on-orbit sources, or independent temperature estimates from the relativistic SZ effect. Improvements by a factor of $\sim 2$ on $H_0$ constraints resulting from these updated calibrations are possible on timescales of one to a few years based on in-progress work. In the longer term, on timescales closer to ten or more years, improvements by an order of magnitude are possible, thus providing per cent level constraints on $H_0$ from this method.
\end{enumerate}

\section*{Acknowledgements}

We thank Michael Zemcov for useful discussions regarding the {\it Herschel}-SPIRE SZ effect data.
JW was supported by a Robert L. Blinkenberg Caltech Summer Undergraduate Research Fellowship and the Kavli Institute for Particle Astrophysics and Cosmology.
We acknowledge support from the U.S. Department of Energy under contract number DE-AC02-76SF00515, and and from the National Aeronautics and Space Administration under Grant No.\ 80NSSC18K0920 issued through the ROSES 2017 Astrophysics Data Analysis Program.

\section*{Data Availability}

The {\it Chandra} data underlying this article are available from the Chandra Data Archive\footnote{\url{https://cxc.cfa.harvard.edu/cda/}}.
The {\it Planck} and Bolocam data products can be obtained from the NASA/IPAC Infrared Science Archive\footnote{\url{https://irsa.ipac.caltech.edu/}} (footnotes 3--4 provide direct URLs).

\bibliographystyle{mnras-morelinks}
\bibliography{bibliography}

\begin{thebibliography}{}
\makeatletter
\relax
\def\mn@urlcharsother{\let\do\@makeother \do\$\do\&\do\#\do\^\do\_\do\%\do\~}
\def\mn@doi{\begingroup\mn@urlcharsother \@ifnextchar [ {\mn@doi@}
  {\mn@doi@[]}}
\def\mn@doi@[#1]#2{\def\@tempa{#1}\ifx\@tempa\@empty \href
  {http://dx.doi.org/#2} {doi:#2}\else \href {http://dx.doi.org/#2} {#1}\fi
  \endgroup}
\def\mn@eprint#1#2{\mn@eprint@#1:#2::\@nil}
\def\mn@eprint@arXiv#1{\href {http://arxiv.org/abs/#1} {{\tt arXiv:#1}}}
\def\mn@eprint@dblp#1{\href {http://dblp.uni-trier.de/rec/bibtex/#1.xml}
  {dblp:#1}}
\def\mn@eprint@#1:#2:#3:#4\@nil{\def\@tempa {#1}\def\@tempb {#2}\def\@tempc
  {#3}\ifx \@tempc \@empty \let \@tempc \@tempb \let \@tempb \@tempa \fi \ifx
  \@tempb \@empty \def\@tempb {arXiv}\fi \@ifundefined
  {mn@eprint@\@tempb}{\@tempb:\@tempc}{\expandafter \expandafter \csname
  mn@eprint@\@tempb\endcsname \expandafter{\@tempc}}}

\bibitem[\protect\citeauthoryear{{Abbott} et~al.,}{{Abbott}
  et~al.}{2017}]{Abbott2017}
{Abbott} B.~P.  et~al., 2017, \mn@doi [\nat] {10.1038/nature24471}, \href
  {https://ui.adsabs.harvard.edu/abs/2017Natur.551...85A} {551, 85}

\bibitem[\protect\citeauthoryear{{Addison}, {Watts}, {Bennett}, {Halpern},
  {Hinshaw}  \& {Weiland}}{{Addison} et~al.}{2018}]{Addison2018}
{Addison} G.~E.,  {Watts} D.~J.,  {Bennett} C.~L.,  {Halpern} M.,  {Hinshaw}
  G.,   {Weiland} J.~L.,  2018, \mn@doi [\apj] {10.3847/1538-4357/aaa1ed},
  \href {https://ui.adsabs.harvard.edu/abs/2018ApJ...853..119A} {853, 119}

\bibitem[\protect\citeauthoryear{{Allen}, {Mantz}, {Morris}, {Applegate},
  {Kelly}, {von der Linden}, {Rapetti}  \& {Schmidt}}{{Allen}
  et~al.}{2013}]{Allen2013}
{Allen} S.~W.,  {Mantz} A.~B.,  {Morris} R.~G.,  {Applegate} D.~E.,  {Kelly}
  P.~L.,  {von der Linden} A.,  {Rapetti} D.~A.,   {Schmidt} R.~W.,  2013,
  preprint, \href {https://ui.adsabs.harvard.edu/abs/2013arXiv1307.8152A}
  {arXiv:1307.8152}

\bibitem[\protect\citeauthoryear{{Allison}, {Sadler}  \& {Meekin}}{{Allison}
  et~al.}{2014}]{A2029photometry2}
{Allison} J.~R.,  {Sadler} E.~M.,   {Meekin} A.~M.,  2014, \mn@doi [\mnras]
  {10.1093/mnras/stu289}, \href
  {https://ui.adsabs.harvard.edu/abs/2014MNRAS.440..696A} {440, 696}

\bibitem[\protect\citeauthoryear{{Ansarifard} et~al.,}{{Ansarifard}
  et~al.}{2020}]{Ansarifard2020}
{Ansarifard} S.  et~al., 2020, \mn@doi [\aap] {10.1051/0004-6361/201936742},
  \href {https://ui.adsabs.harvard.edu/abs/2020A&A...634A.113A} {634, A113}

\bibitem[\protect\citeauthoryear{{Applegate} et~al.,}{{Applegate}
  et~al.}{2014}]{Applegate2014}
{Applegate} D.~E.  et~al., 2014, \mn@doi [\mnras] {10.1093/mnras/stt2129},
  \href {https://ui.adsabs.harvard.edu/abs/2014MNRAS.439...48A} {439, 48}

\bibitem[\protect\citeauthoryear{{Applegate} et~al.,}{{Applegate}
  et~al.}{2016}]{Applegate2016}
{Applegate} D.~E.  et~al., 2016, \mn@doi [\mnras] {10.1093/mnras/stw005}, \href
  {https://ui.adsabs.harvard.edu/abs/2016MNRAS.457.1522A} {457, 1522}

\bibitem[\protect\citeauthoryear{{Arnaud}, {Pratt}, {Piffaretti},
  {B{\"o}hringer}, {Croston}  \& {Pointecouteau}}{{Arnaud}
  et~al.}{2010}]{Arnaud2010}
{Arnaud} M.,  {Pratt} G.~W.,  {Piffaretti} R.,  {B{\"o}hringer} H.,  {Croston}
  J.~H.,   {Pointecouteau} E.,  2010, \mn@doi [\aap]
  {10.1051/0004-6361/200913416}, \href
  {https://ui.adsabs.harvard.edu/abs/2010A&A...517A..92A} {517, A92}

\bibitem[\protect\citeauthoryear{{Aubourg} et~al.,}{{Aubourg}
  et~al.}{2015}]{Aubourg2015}
{Aubourg} {\'E}.  et~al., 2015, \mn@doi [\prd] {10.1103/PhysRevD.92.123516},
  \href {https://ui.adsabs.harvard.edu/abs/2015PhRvD..92l3516A} {92, 123516}

\bibitem[\protect\citeauthoryear{{Battaglia}, {Bond}, {Pfrommer}  \&
  {Sievers}}{{Battaglia} et~al.}{2015}]{Battaglia2015}
{Battaglia} N.,  {Bond} J.~R.,  {Pfrommer} C.,   {Sievers} J.~L.,  2015,
  \mn@doi [\apj] {10.1088/0004-637X/806/1/43}, \href
  {https://ui.adsabs.harvard.edu/abs/2015ApJ...806...43B} {806, 43}

\bibitem[\protect\citeauthoryear{{Becker}, {White}  \& {Edwards}}{{Becker}
  et~al.}{1991}]{A2029photometry4}
{Becker} R.~H.,  {White} R.~L.,   {Edwards} A.~L.,  1991, \mn@doi [\apjs]
  {10.1086/191529}, \href
  {https://ui.adsabs.harvard.edu/abs/1991ApJS...75....1B} {75, 1}

\bibitem[\protect\citeauthoryear{{Berlok}, {Quataert}, {Pessah}  \&
  {Pfrommer}}{{Berlok} et~al.}{2020}]{Berlok2007.00018}
{Berlok} T.,  {Quataert} E.,  {Pessah} M.~E.,   {Pfrommer} C.,  2020, preprint,
  \href {https://ui.adsabs.harvard.edu/abs/2020arXiv200700018B}
  {arXiv:2007.00018}

\bibitem[\protect\citeauthoryear{{Bertincourt} et~al.,}{{Bertincourt}
  et~al.}{2016}]{Bertincourt2016}
{Bertincourt} B.  et~al., 2016, \mn@doi [\aap] {10.1051/0004-6361/201527313},
  \href {https://ui.adsabs.harvard.edu/abs/2016A&A...588A.107B} {588, A107}

\bibitem[\protect\citeauthoryear{{Biffi}, {Sembolini}, {De Petris},
  {Valdarnini}, {Yepes}  \& {Gottl{\"o}ber}}{{Biffi} et~al.}{2014}]{Biffi2014}
{Biffi} V.,  {Sembolini} F.,  {De Petris} M.,  {Valdarnini} R.,  {Yepes} G.,
  {Gottl{\"o}ber} S.,  2014, \mn@doi [\mnras] {10.1093/mnras/stu018}, \href
  {https://ui.adsabs.harvard.edu/abs/2014MNRAS.439..588B} {439, 588}

\bibitem[\protect\citeauthoryear{{Birkinshaw}}{{Birkinshaw}}{1979}]{Birkinshaw1979}
{Birkinshaw} M.,  1979, \mn@doi [\mnras] {10.1093/mnras/187.4.847}, \href
  {https://ui.adsabs.harvard.edu/abs/1979MNRAS.187..847B} {187, 847}

\bibitem[\protect\citeauthoryear{{Birkinshaw}, {Hughes}  \&
  {Arnaud}}{{Birkinshaw} et~al.}{1991}]{Birkinshaw1991}
{Birkinshaw} M.,  {Hughes} J.~P.,   {Arnaud} K.~A.,  1991, \mn@doi [\apj]
  {10.1086/170522}, \href
  {https://ui.adsabs.harvard.edu/abs/1991ApJ...379..466B} {379, 466}

\bibitem[\protect\citeauthoryear{{Birrer} et~al.,}{{Birrer}
  et~al.}{2020}]{Birrer2020}
{Birrer} S.  et~al., 2020, \mn@doi [\aap] {10.1051/0004-6361/202038861}, \href
  {https://ui.adsabs.harvard.edu/abs/2020A&A...643A.165B} {643, A165}

\bibitem[\protect\citeauthoryear{{Bonamente}, {Joy}, {LaRoque}, {Carlstrom},
  {Reese}  \& {Dawson}}{{Bonamente} et~al.}{2006}]{Bonamente2006}
{Bonamente} M.,  {Joy} M.~K.,  {LaRoque} S.~J.,  {Carlstrom} J.~E.,  {Reese}
  E.~D.,   {Dawson} K.~S.,  2006, \mn@doi [\apj] {10.1086/505291}, \href
  {https://ui.adsabs.harvard.edu/abs/2006ApJ...647...25B} {647, 25}

\bibitem[\protect\citeauthoryear{{Bonamigo}, {Despali}, {Limousin}, {Angulo},
  {Giocoli}  \& {Soucail}}{{Bonamigo} et~al.}{2015}]{Bonamigo2015}
{Bonamigo} M.,  {Despali} G.,  {Limousin} M.,  {Angulo} R.,  {Giocoli} C.,
  {Soucail} G.,  2015, \mn@doi [\mnras] {10.1093/mnras/stv417}, \href
  {https://ui.adsabs.harvard.edu/abs/2015MNRAS.449.3171B} {449, 3171}

\bibitem[\protect\citeauthoryear{{Carilli} \& {Taylor}}{{Carilli} \&
  {Taylor}}{2002}]{Carilli0110655}
{Carilli} C.~L.,  {Taylor} G.~B.,  2002, \mn@doi [\araa]
  {10.1146/annurev.astro.40.060401.093852}, \href
  {https://ui.adsabs.harvard.edu/abs/2002ARA&A..40..319C} {40, 319}

\bibitem[\protect\citeauthoryear{{Chluba}, {Nagai}, {Sazonov}  \&
  {Nelson}}{{Chluba} et~al.}{2012}]{Chluba2012}
{Chluba} J.,  {Nagai} D.,  {Sazonov} S.,   {Nelson} K.,  2012, \mn@doi [\mnras]
  {10.1111/j.1365-2966.2012.21741.x}, \href
  {https://ui.adsabs.harvard.edu/abs/2012MNRAS.426..510C} {426, 510}

\bibitem[\protect\citeauthoryear{{Chluba}, {Switzer}, {Nelson}  \&
  {Nagai}}{{Chluba} et~al.}{2013}]{Chluba2013}
{Chluba} J.,  {Switzer} E.,  {Nelson} K.,   {Nagai} D.,  2013, \mn@doi [\mnras]
  {10.1093/mnras/stt110}, \href
  {https://ui.adsabs.harvard.edu/abs/2013MNRAS.430.3054C} {430, 3054}

\bibitem[\protect\citeauthoryear{{Choi} et~al.,}{{Choi}
  et~al.}{2020}]{Choi2020}
{Choi} S.~K.  et~al., 2020, preprint, \href
  {https://ui.adsabs.harvard.edu/abs/2020arXiv200707289C} {arXiv:2007.07289}

\bibitem[\protect\citeauthoryear{{Coble} et~al.,}{{Coble}
  et~al.}{2007}]{A478photometry1}
{Coble} K.  et~al., 2007, \mn@doi [\aj] {10.1086/519973}, \href
  {https://ui.adsabs.harvard.edu/abs/2007AJ....134..897C} {134, 897}

\bibitem[\protect\citeauthoryear{{Condon}, {Cotton}, {Greisen}, {Yin},
  {Perley}, {Taylor}  \& {Broderick}}{{Condon} et~al.}{1998}]{A2029photometry1}
{Condon} J.~J.,  {Cotton} W.~D.,  {Greisen} E.~W.,  {Yin} Q.~F.,  {Perley}
  R.~A.,  {Taylor} G.~B.,   {Broderick} J.~J.,  1998, \mn@doi [\aj]
  {10.1086/300337}, \href
  {https://ui.adsabs.harvard.edu/abs/1998AJ....115.1693C} {115, 1693}

\bibitem[\protect\citeauthoryear{{Drake}, {Pfrommer}, {Reynolds}, {Ruszkowski},
  {Swisdak}, {Einarsson}, {Hassam}  \& {Roberg-Clark}}{{Drake}
  et~al.}{2020}]{Drake2007.07931}
{Drake} J.~F.,  {Pfrommer} C.,  {Reynolds} C.~S.,  {Ruszkowski} M.,  {Swisdak}
  M.,  {Einarsson} A.,  {Hassam} A.~B.,   {Roberg-Clark} G.~T.,  2020,
  preprint, \href {https://ui.adsabs.harvard.edu/abs/2020arXiv200707931D}
  {arXiv:2007.07931}

\bibitem[\protect\citeauthoryear{{Eckert}, {Roncarelli}, {Ettori}, {Molendi},
  {Vazza}, {Gastaldello}  \& {Rossetti}}{{Eckert} et~al.}{2015}]{Eckert2015}
{Eckert} D.,  {Roncarelli} M.,  {Ettori} S.,  {Molendi} S.,  {Vazza} F.,
  {Gastaldello} F.,   {Rossetti} M.,  2015, \mn@doi [\mnras]
  {10.1093/mnras/stu2590}, \href
  {https://ui.adsabs.harvard.edu/abs/2015MNRAS.447.2198E} {447, 2198}

\bibitem[\protect\citeauthoryear{{Egami} et~al.,}{{Egami}
  et~al.}{2010}]{Egami2010}
{Egami} E.  et~al., 2010, \mn@doi [\aap] {10.1051/0004-6361/201014696}, \href
  {https://ui.adsabs.harvard.edu/abs/2010A&A...518L..12E} {518, L12}

\bibitem[\protect\citeauthoryear{{Ettori} \& {Fabian}}{{Ettori} \&
  {Fabian}}{2006}]{Ettori0603383}
{Ettori} S.,  {Fabian} A.~C.,  2006, \mn@doi [\mnras]
  {10.1111/j.1745-3933.2006.00170.x}, \href
  {https://ui.adsabs.harvard.edu/abs/2006MNRAS.369L..42E} {369, L42}

\bibitem[\protect\citeauthoryear{{Fabian}}{{Fabian}}{2012}]{Fabian1204.4114}
{Fabian} A.~C.,  2012, \mn@doi [\araa] {10.1146/annurev-astro-081811-125521},
  \href {https://ui.adsabs.harvard.edu/abs/2012ARA&A..50..455F} {50, 455}

\bibitem[\protect\citeauthoryear{{Foreman-Mackey}, {Hogg}, {Lang}  \&
  {Goodman}}{{Foreman-Mackey} et~al.}{2013}]{Foreman-Mackey2013}
{Foreman-Mackey} D.,  {Hogg} D.~W.,  {Lang} D.,   {Goodman} J.,  2013, \mn@doi
  [\pasp] {10.1086/670067}, \href
  {https://ui.adsabs.harvard.edu/abs/2013PASP..125..306F} {125, 306}

\bibitem[\protect\citeauthoryear{{Freedman} et~al.,}{{Freedman}
  et~al.}{2020}]{Freedman2020}
{Freedman} W.~L.  et~al., 2020, \mn@doi [\apj] {10.3847/1538-4357/ab7339},
  \href {https://ui.adsabs.harvard.edu/abs/2020ApJ...891...57F} {891, 57}

\bibitem[\protect\citeauthoryear{{Gregory} \& {Condon}}{{Gregory} \&
  {Condon}}{1991}]{A2029photometry5}
{Gregory} P.~C.,  {Condon} J.~J.,  1991, \mn@doi [\apjs] {10.1086/191559},
  \href {https://ui.adsabs.harvard.edu/abs/1991ApJS...75.1011G} {75, 1011}

\bibitem[\protect\citeauthoryear{{Griffith}, {Wright}, {Burke}  \&
  {Ekers}}{{Griffith} et~al.}{1994}]{PKS0745photometry2}
{Griffith} M.~R.,  {Wright} A.~E.,  {Burke} B.~F.,   {Ekers} R.~D.,  1994,
  \mn@doi [\apjs] {10.1086/191863}, \href
  {https://ui.adsabs.harvard.edu/abs/1994ApJS...90..179G} {90, 179}

\bibitem[\protect\citeauthoryear{{Griffith}, {Wright}, {Burke}  \&
  {Ekers}}{{Griffith} et~al.}{1995}]{A2029photometry6}
{Griffith} M.~R.,  {Wright} A.~E.,  {Burke} B.~F.,   {Ekers} R.~D.,  1995,
  \mn@doi [\apjs] {10.1086/192146}, \href
  {https://ui.adsabs.harvard.edu/abs/1995ApJS...97..347G} {97, 347}

\bibitem[\protect\citeauthoryear{{Hitomi Collaboration},}{{Hitomi
  Collaboration}}{2016}]{Hitomi1607.04487}
{Hitomi Collaboration}, 2016, \mn@doi [\nat] {10.1038/nature18627}, \href
  {https://ui.adsabs.harvard.edu/abs/2016Natur.535..117H} {535, 117}

\bibitem[\protect\citeauthoryear{{Hou} et~al.,}{{Hou} et~al.}{2018}]{Hou2018}
{Hou} Z.  et~al., 2018, \mn@doi [\apj] {10.3847/1538-4357/aaa3ef}, \href
  {https://ui.adsabs.harvard.edu/abs/2018ApJ...853....3H} {853, 3}

\bibitem[\protect\citeauthoryear{{Jahoda} et~al.,}{{Jahoda}
  et~al.}{2019}]{Jahoda2019}
{Jahoda} K.  et~al., 2019, in Bulletin of the American Astronomical Society.
  p.~174 (\mn@eprint {arXiv} {1907.12677})

\bibitem[\protect\citeauthoryear{{Jetha}, {Hardcastle}  \& {Sakelliou}}{{Jetha}
  et~al.}{2006}]{A2029photometry8}
{Jetha} N.~N.,  {Hardcastle} M.~J.,   {Sakelliou} I.,  2006, \mn@doi [\mnras]
  {10.1111/j.1365-2966.2006.10155.x}, \href
  {https://ui.adsabs.harvard.edu/abs/2006MNRAS.368..609J} {368, 609}

\bibitem[\protect\citeauthoryear{{Jing} \& {Suto}}{{Jing} \&
  {Suto}}{2002}]{Jing2002}
{Jing} Y.~P.,  {Suto} Y.,  2002, \mn@doi [\apj] {10.1086/341065}, \href
  {https://ui.adsabs.harvard.edu/abs/2002ApJ...574..538J} {574, 538}

\bibitem[\protect\citeauthoryear{{Jones} et~al.,}{{Jones}
  et~al.}{2005}]{Jones2005}
{Jones} M.~E.  et~al., 2005, \mn@doi [\mnras]
  {10.1111/j.1365-2966.2005.08626.x}, \href
  {https://ui.adsabs.harvard.edu/abs/2005MNRAS.357..518J} {357, 518}

\bibitem[\protect\citeauthoryear{{Kalberla}, {Burton}, {Hartmann}, {Arnal},
  {Bajaja}, {Morras}  \& {P{\"o}ppel}}{{Kalberla} et~al.}{2005}]{Kalberla2005}
{Kalberla} P.~M.~W.,  {Burton} W.~B.,  {Hartmann} D.,  {Arnal} E.~M.,  {Bajaja}
  E.,  {Morras} R.,   {P{\"o}ppel} W.~G.~L.,  2005, \mn@doi [\aap]
  {10.1051/0004-6361:20041864}, \href
  {https://ui.adsabs.harvard.edu/abs/2005A&A...440..775K} {440, 775}

\bibitem[\protect\citeauthoryear{{Kawahara}, {Kitayama}, {Sasaki}  \&
  {Suto}}{{Kawahara} et~al.}{2008}]{Kawahara2008}
{Kawahara} H.,  {Kitayama} T.,  {Sasaki} S.,   {Suto} Y.,  2008, \mn@doi [\apj]
  {10.1086/524132}, \href
  {https://ui.adsabs.harvard.edu/abs/2008ApJ...674...11K} {674, 11}

\bibitem[\protect\citeauthoryear{{Kellermann} \& {Pauliny-Toth}}{{Kellermann}
  \& {Pauliny-Toth}}{1973}]{PKS0745photometry3}
{Kellermann} K.~I.,  {Pauliny-Toth} I.~I.~K.,  1973, \mn@doi [\aj]
  {10.1086/111489}, \href
  {https://ui.adsabs.harvard.edu/abs/1973AJ.....78..828K} {78, 828}

\bibitem[\protect\citeauthoryear{{Kim} et~al.,}{{Kim}
  et~al.}{2013}]{Kim1309.5382}
{Kim} A.  et~al., 2013, arXiv:1309.5382, \href
  {http://adsabs.harvard.edu/abs/2013arXiv1309.5382K} {}

\bibitem[\protect\citeauthoryear{{Klaassen} et~al.,}{{Klaassen}
  et~al.}{2020}]{Klaassen2020}
{Klaassen} P.~D.  et~al., 2020, preprint, \href
  {https://ui.adsabs.harvard.edu/abs/2020arXiv201107974K} {arXiv:2011.07974}

\bibitem[\protect\citeauthoryear{{Kozmanyan}, {Bourdin}, {Mazzotta}, {Rasia}
  \& {Sereno}}{{Kozmanyan} et~al.}{2019}]{Kozmanyan2019}
{Kozmanyan} A.,  {Bourdin} H.,  {Mazzotta} P.,  {Rasia} E.,   {Sereno} M.,
  2019, \mn@doi [\aap] {10.1051/0004-6361/201833879}, \href
  {https://ui.adsabs.harvard.edu/abs/2019A&A...621A..34K} {621, A34}

\bibitem[\protect\citeauthoryear{{Lau}, {Nagai}, {Avestruz}, {Nelson}  \&
  {Vikhlinin}}{{Lau} et~al.}{2015}]{Lau2015}
{Lau} E.~T.,  {Nagai} D.,  {Avestruz} C.,  {Nelson} K.,   {Vikhlinin} A.,
  2015, \mn@doi [\apj] {10.1088/0004-637X/806/1/68}, \href
  {https://ui.adsabs.harvard.edu/abs/2015ApJ...806...68L} {806, 68}

\bibitem[\protect\citeauthoryear{{Lau}, {Gaspari}, {Nagai}  \& {Coppi}}{{Lau}
  et~al.}{2017}]{Lau1705.06280}
{Lau} E.~T.,  {Gaspari} M.,  {Nagai} D.,   {Coppi} P.,  2017, \mn@doi [\apj]
  {10.3847/1538-4357/aa8c00}, \href
  {https://ui.adsabs.harvard.edu/abs/2017ApJ...849...54L} {849, 54}

\bibitem[\protect\citeauthoryear{{Lau}, {Hearin}, {Nagai}  \&
  {Cappelluti}}{{Lau} et~al.}{2020}]{Lau2020}
{Lau} E.~T.,  {Hearin} A.~P.,  {Nagai} D.,   {Cappelluti} N.,  2020, preprint,
  \href {https://ui.adsabs.harvard.edu/abs/2020arXiv200609420L}
  {arXiv:2006.09420}

\bibitem[\protect\citeauthoryear{Lenth}{Lenth}{1989}]{Lenth2347693}
Lenth R.~V.,  1989, \mn@doi [Journal of the Royal Statistical Society. Series C
  (Applied Statistics)] {10.2307/2347693}, \href
  {http://www.jstor.org/stable/2347693} {38, 185}

\bibitem[\protect\citeauthoryear{{Lin}, {Partridge}, {Pober}, {Bouchefry},
  {Burke}, {Klein}, {Coish}  \& {Huffenberger}}{{Lin}
  et~al.}{2009}]{A2029photometry7}
{Lin} Y.-T.,  {Partridge} B.,  {Pober} J.~C.,  {Bouchefry} K.~E.,  {Burke} S.,
  {Klein} J.~N.,  {Coish} J.~W.,   {Huffenberger} K.~M.,  2009, \mn@doi [\apj]
  {10.1088/0004-637X/694/2/992}, \href
  {https://ui.adsabs.harvard.edu/abs/2009ApJ...694..992L} {694, 992}

\bibitem[\protect\citeauthoryear{Livio \& Riess}{Livio \&
  Riess}{2013}]{Livio2013}
Livio M.,  Riess A.,  2013, \mn@doi [Physics Today] {10.1063/PT.3.2148}, 66, 41

\bibitem[\protect\citeauthoryear{{Louis} et~al.,}{{Louis}
  et~al.}{2014}]{Louis2014}
{Louis} T.  et~al., 2014, \mn@doi [\jcap] {10.1088/1475-7516/2014/07/016},
  \href {https://ui.adsabs.harvard.edu/abs/2014JCAP...07..016L} {2014, 016}

\bibitem[\protect\citeauthoryear{{Mantz}, {Allen}, {Morris}, {Rapetti},
  {Applegate}, {Kelly}, {von der Linden}  \& {Schmidt}}{{Mantz}
  et~al.}{2014}]{Mantz1402.6212}
{Mantz} A.~B.,  {Allen} S.~W.,  {Morris} R.~G.,  {Rapetti} D.~A.,  {Applegate}
  D.~E.,  {Kelly} P.~L.,  {von der Linden} A.,   {Schmidt} R.~W.,  2014,
  \mn@doi [\mnras] {10.1093/mnras/stu368}, \href
  {https://ui.adsabs.harvard.edu/abs/2014MNRAS.440.2077M} {440, 2077}

\bibitem[\protect\citeauthoryear{{Mantz}, {Allen}, {Morris}, {Schmidt}, {von
  der Linden}  \& {Urban}}{{Mantz} et~al.}{2015}]{Mantz1502.06020}
{Mantz} A.~B.,  {Allen} S.~W.,  {Morris} R.~G.,  {Schmidt} R.~W.,  {von der
  Linden} A.,   {Urban} O.,  2015, \mn@doi [\mnras] {10.1093/mnras/stv219},
  \href {https://ui.adsabs.harvard.edu/abs/2015MNRAS.449..199M} {449, 199}

\bibitem[\protect\citeauthoryear{{Mantz}, {Allen}, {Morris}  \&
  {Schmidt}}{{Mantz} et~al.}{2016}]{Mantz1509.01322}
{Mantz} A.~B.,  {Allen} S.~W.,  {Morris} R.~G.,   {Schmidt} R.~W.,  2016,
  \mn@doi [\mnras] {10.1093/mnras/stv2899}, \href
  {https://ui.adsabs.harvard.edu/abs/2016MNRAS.456.4020M} {456, 4020}

\bibitem[\protect\citeauthoryear{{Markwardt}}{{Markwardt}}{2009}]{Markwardt2009}
{Markwardt} C.~B.,  2009, in {Bohlender} D.~A.,  {Durand} D.,   {Dowler} P.,
  eds,  Astronomical Society of the Pacific Conference Series Vol. 411,
  Astronomical Data Analysis Software and Systems XVIII. p.~251 (\mn@eprint
  {arXiv} {0902.2850})

\bibitem[\protect\citeauthoryear{{Mason}, {Myers}  \& {Readhead}}{{Mason}
  et~al.}{2001}]{Mason2001}
{Mason} B.~S.,  {Myers} S.~T.,   {Readhead} A.~C.~S.,  2001, \mn@doi [\apjl]
  {10.1086/321737}, \href
  {https://ui.adsabs.harvard.edu/abs/2001ApJ...555L..11M} {555, L11}

\bibitem[\protect\citeauthoryear{{McNamara} \& {Nulsen}}{{McNamara} \&
  {Nulsen}}{2007}]{McNamara2007}
{McNamara} B.~R.,  {Nulsen} P.~E.~J.,  2007, \mn@doi [\araa]
  {10.1146/annurev.astro.45.051806.110625}, \href
  {https://ui.adsabs.harvard.edu/abs/2007ARA&A..45..117M} {45, 117}

\bibitem[\protect\citeauthoryear{{Miville-Desch{\^e}nes} \&
  {Lagache}}{{Miville-Desch{\^e}nes} \& {Lagache}}{2005}]{Miville-Desch2005}
{Miville-Desch{\^e}nes} M.-A.,  {Lagache} G.,  2005, \mn@doi [\apjs]
  {10.1086/427938}, \href
  {https://ui.adsabs.harvard.edu/abs/2005ApJS..157..302M} {157, 302}

\bibitem[\protect\citeauthoryear{{Morandi}, {Nagai}  \& {Cui}}{{Morandi}
  et~al.}{2013}]{Morandi2013}
{Morandi} A.,  {Nagai} D.,   {Cui} W.,  2013, \mn@doi [\mnras]
  {10.1093/mnras/stt252}, \href
  {https://ui.adsabs.harvard.edu/abs/2013MNRAS.431.1240M} {431, 1240}

\bibitem[\protect\citeauthoryear{{Mueller}, {de Bernardis}, {Bean}  \&
  {Niemack}}{{Mueller} et~al.}{2015}]{Mueller2015}
{Mueller} E.-M.,  {de Bernardis} F.,  {Bean} R.,   {Niemack} M.~D.,  2015,
  \mn@doi [\apj] {10.1088/0004-637X/808/1/47}, \href
  {https://ui.adsabs.harvard.edu/abs/2015ApJ...808...47M} {808, 47}

\bibitem[\protect\citeauthoryear{{Nagai}, {Vikhlinin}  \& {Kravtsov}}{{Nagai}
  et~al.}{2007}]{Nagai2007}
{Nagai} D.,  {Vikhlinin} A.,   {Kravtsov} A.~V.,  2007, \mn@doi [\apj]
  {10.1086/509868}, \href
  {https://ui.adsabs.harvard.edu/abs/2007ApJ...655...98N} {655, 98}

\bibitem[\protect\citeauthoryear{{Nevalainen}, {David}  \&
  {Guainazzi}}{{Nevalainen} et~al.}{2010}]{Nevalainen2010}
{Nevalainen} J.,  {David} L.,   {Guainazzi} M.,  2010, \mn@doi [\aap]
  {10.1051/0004-6361/201015176}, \href
  {https://ui.adsabs.harvard.edu/abs/2010A&A...523A..22N} {523, A22}

\bibitem[\protect\citeauthoryear{{Oliver} et~al.,}{{Oliver}
  et~al.}{2012}]{Oliver2012}
{Oliver} S.~J.  et~al., 2012, \mn@doi [\mnras]
  {10.1111/j.1365-2966.2012.20912.x}, \href
  {https://ui.adsabs.harvard.edu/abs/2012MNRAS.424.1614O} {424, 1614}

\bibitem[\protect\citeauthoryear{{Parrish}, {Quataert}  \& {Sharma}}{{Parrish}
  et~al.}{2009}]{Parrish0905.4500}
{Parrish} I.~J.,  {Quataert} E.,   {Sharma} P.,  2009, \mn@doi [\apj]
  {10.1088/0004-637X/703/1/96}, \href
  {https://ui.adsabs.harvard.edu/abs/2009ApJ...703...96P} {703, 96}

\bibitem[\protect\citeauthoryear{{Pesce} et~al.,}{{Pesce}
  et~al.}{2020}]{Pesce2020}
{Pesce} D.~W.  et~al., 2020, \mn@doi [\apjl] {10.3847/2041-8213/ab75f0}, \href
  {https://ui.adsabs.harvard.edu/abs/2020ApJ...891L...1P} {891, L1}

\bibitem[\protect\citeauthoryear{{Piffaretti}, {Arnaud}, {Pratt},
  {Pointecouteau}  \& {Melin}}{{Piffaretti} et~al.}{2011}]{Piffaretti2011}
{Piffaretti} R.,  {Arnaud} M.,  {Pratt} G.~W.,  {Pointecouteau} E.,   {Melin}
  J.~B.,  2011, \mn@doi [\aap] {10.1051/0004-6361/201015377}, \href
  {https://ui.adsabs.harvard.edu/abs/2011A&A...534A.109P} {534, A109}

\bibitem[\protect\citeauthoryear{{Planck Collaboration},}{{Planck
  Collaboration}}{2016a}]{Planck2016_VIII}
{Planck Collaboration}, 2016a, \mn@doi [\aap] {10.1051/0004-6361/201525820},
  \href {https://ui.adsabs.harvard.edu/abs/2016A&A...594A...8P} {594, A8}

\bibitem[\protect\citeauthoryear{{Planck Collaboration},}{{Planck
  Collaboration}}{2016b}]{Planck2016_XXII}
{Planck Collaboration}, 2016b, \mn@doi [\aap] {10.1051/0004-6361/201525826},
  \href {https://ui.adsabs.harvard.edu/abs/2016A&A...594A..22P} {594, A22}

\bibitem[\protect\citeauthoryear{{Planck Collaboration},}{{Planck
  Collaboration}}{2017}]{Planck2017_LII}
{Planck Collaboration}, 2017, \mn@doi [\aap] {10.1051/0004-6361/201630311},
  \href {https://ui.adsabs.harvard.edu/abs/2017A&A...607A.122P} {607, A122}

\bibitem[\protect\citeauthoryear{{Planck Collaboration},}{{Planck
  Collaboration}}{2020}]{Planck2018_VI}
{Planck Collaboration}, 2020, \mn@doi [\aap] {10.1051/0004-6361/201833910},
  \href {https://ui.adsabs.harvard.edu/abs/2020A&A...641A...6P} {641, A6}

\bibitem[\protect\citeauthoryear{{Planelles} et~al.,}{{Planelles}
  et~al.}{2017}]{Planelles2017}
{Planelles} S.  et~al., 2017, \mn@doi [\mnras] {10.1093/mnras/stx318}, \href
  {https://ui.adsabs.harvard.edu/abs/2017MNRAS.467.3827P} {467, 3827}

\bibitem[\protect\citeauthoryear{{Poulin}, {Smith}, {Karwal}  \&
  {Kamionkowski}}{{Poulin} et~al.}{2019}]{Poulin2019}
{Poulin} V.,  {Smith} T.~L.,  {Karwal} T.,   {Kamionkowski} M.,  2019, \mn@doi
  [\prl] {10.1103/PhysRevLett.122.221301}, \href
  {https://ui.adsabs.harvard.edu/abs/2019PhRvL.122v1301P} {122, 221301}

\bibitem[\protect\citeauthoryear{{Reese} et~al.,}{{Reese}
  et~al.}{2000}]{Reese2000}
{Reese} E.~D.  et~al., 2000, \mn@doi [\apj] {10.1086/308662}, \href
  {https://ui.adsabs.harvard.edu/abs/2000ApJ...533...38R} {533, 38}

\bibitem[\protect\citeauthoryear{{Reese}, {Carlstrom}, {Joy}, {Mohr}, {Grego}
  \& {Holzapfel}}{{Reese} et~al.}{2002}]{Reese2002}
{Reese} E.~D.,  {Carlstrom} J.~E.,  {Joy} M.,  {Mohr} J.~J.,  {Grego} L.,
  {Holzapfel} W.~L.,  2002, \mn@doi [\apj] {10.1086/344137}, \href
  {https://ui.adsabs.harvard.edu/abs/2002ApJ...581...53R} {581, 53}

\bibitem[\protect\citeauthoryear{{Riess}, {Casertano}, {Yuan}, {Macri}  \&
  {Scolnic}}{{Riess} et~al.}{2019}]{Riess2019}
{Riess} A.~G.,  {Casertano} S.,  {Yuan} W.,  {Macri} L.~M.,   {Scolnic} D.,
  2019, \mn@doi [\apj] {10.3847/1538-4357/ab1422}, \href
  {https://ui.adsabs.harvard.edu/abs/2019ApJ...876...85R} {876, 85}

\bibitem[\protect\citeauthoryear{{Roncarelli}, {Ettori}, {Borgani}, {Dolag},
  {Fabjan}  \& {Moscardini}}{{Roncarelli} et~al.}{2013}]{Roncarelli2013}
{Roncarelli} M.,  {Ettori} S.,  {Borgani} S.,  {Dolag} K.,  {Fabjan} D.,
  {Moscardini} L.,  2013, \mn@doi [\mnras] {10.1093/mnras/stt654}, \href
  {https://ui.adsabs.harvard.edu/abs/2013MNRAS.432.3030R} {432, 3030}

\bibitem[\protect\citeauthoryear{{Rozo}, {Rykoff}, {Bartlett}  \&
  {Evrard}}{{Rozo} et~al.}{2014}]{Rozo2014}
{Rozo} E.,  {Rykoff} E.~S.,  {Bartlett} J.~G.,   {Evrard} A.,  2014, \mn@doi
  [\mnras] {10.1093/mnras/stt2091}, \href
  {https://ui.adsabs.harvard.edu/abs/2014MNRAS.438...49R} {438, 49}

\bibitem[\protect\citeauthoryear{{Sayers} et~al.,}{{Sayers}
  et~al.}{2013a}]{sayers2013radio}
{Sayers} J.  et~al., 2013a, \mn@doi [\apj] {10.1088/0004-637X/764/2/152}, \href
  {https://ui.adsabs.harvard.edu/abs/2013ApJ...764..152S} {764, 152}

\bibitem[\protect\citeauthoryear{{Sayers} et~al.,}{{Sayers}
  et~al.}{2013b}]{Sayers2013}
{Sayers} J.  et~al., 2013b, \mn@doi [\apj] {10.1088/0004-637X/768/2/177}, \href
  {https://ui.adsabs.harvard.edu/abs/2013ApJ...768..177S} {768, 177}

\bibitem[\protect\citeauthoryear{{Sayers} et~al.,}{{Sayers}
  et~al.}{2016}]{Sayers2016}
{Sayers} J.  et~al., 2016, \mn@doi [\apj] {10.3847/0004-637X/832/1/26}, \href
  {https://ui.adsabs.harvard.edu/abs/2016ApJ...832...26S} {832, 26}

\bibitem[\protect\citeauthoryear{{Sayers} et~al.,}{{Sayers}
  et~al.}{2019}]{Sayers2019}
{Sayers} J.  et~al., 2019, \mn@doi [\apj] {10.3847/1538-4357/ab29ef}, \href
  {https://ui.adsabs.harvard.edu/abs/2019ApJ...880...45S} {880, 45}

\bibitem[\protect\citeauthoryear{{Schellenberger}, {Reiprich}, {Lovisari},
  {Nevalainen}  \& {David}}{{Schellenberger} et~al.}{2015}]{Schellenberger2015}
{Schellenberger} G.,  {Reiprich} T.~H.,  {Lovisari} L.,  {Nevalainen} J.,
  {David} L.,  2015, \mn@doi [\aap] {10.1051/0004-6361/201424085}, \href
  {https://ui.adsabs.harvard.edu/abs/2015A&A...575A..30S} {575, A30}

\bibitem[\protect\citeauthoryear{{Schmidt}, Allen  \& Fabian}{{Schmidt}
  et~al.}{2004}]{Schmidt2004}
{Schmidt} R.~W.,  Allen S.~W.,   Fabian A.~C.,  2004, \mn@doi [\mnras]
  {10.1111/j.1365-2966.2004.08032.x}, 352, 1413–1420

\bibitem[\protect\citeauthoryear{{Sheth} \& {Diaferio}}{{Sheth} \&
  {Diaferio}}{2001}]{Sheth2001}
{Sheth} R.~K.,  {Diaferio} A.,  2001, \mn@doi [\mnras]
  {10.1046/j.1365-8711.2001.04202.x}, \href
  {https://ui.adsabs.harvard.edu/abs/2001MNRAS.322..901S} {322, 901}

\bibitem[\protect\citeauthoryear{{Silk} \& {White}}{{Silk} \&
  {White}}{1978}]{Silk1978}
{Silk} J.,  {White} S.~D.~M.,  1978, \mn@doi [\apjl] {10.1086/182841}, \href
  {https://ui.adsabs.harvard.edu/abs/1978ApJ...226L.103S} {226, L103}

\bibitem[\protect\citeauthoryear{{Simionescu} et~al.,}{{Simionescu}
  et~al.}{2011}]{Simionescu2011}
{Simionescu} A.  et~al., 2011, \mn@doi [Science] {10.1126/science.1200331},
  \href {https://ui.adsabs.harvard.edu/abs/2011Sci...331.1576S} {331, 1576}

\bibitem[\protect\citeauthoryear{{Sulkanen}}{{Sulkanen}}{1999}]{Sulkanen1999}
{Sulkanen} M.~E.,  1999, \mn@doi [\apj] {10.1086/307615}, \href
  {https://ui.adsabs.harvard.edu/abs/1999ApJ...522...59S} {522, 59}

\bibitem[\protect\citeauthoryear{{Sunyaev} \& {Zeldovich}}{{Sunyaev} \&
  {Zeldovich}}{1972}]{Sunyaev1972}
{Sunyaev} R.~A.,  {Zeldovich} Y.~B.,  1972, Comments on Astrophysics and Space
  Physics, \href {https://ui.adsabs.harvard.edu/abs/1972CoASP...4..173S} {4,
  173}

\bibitem[\protect\citeauthoryear{{Sunyaev} \& {Zeldovich}}{{Sunyaev} \&
  {Zeldovich}}{1980}]{Sunyaev1980}
{Sunyaev} R.~A.,  {Zeldovich} Y.~B.,  1980, \mn@doi [\mnras]
  {10.1093/mnras/190.3.413}, \href
  {https://ui.adsabs.harvard.edu/abs/1980MNRAS.190..413S} {190, 413}

\bibitem[\protect\citeauthoryear{{Urban} et~al.,}{{Urban}
  et~al.}{2014}]{Urban2014}
{Urban} O.  et~al., 2014, \mn@doi [\mnras] {10.1093/mnras/stt2209}, \href
  {https://ui.adsabs.harvard.edu/abs/2014MNRAS.437.3939U} {437, 3939}

\bibitem[\protect\citeauthoryear{{Wright} \& {Otrupcek}}{{Wright} \&
  {Otrupcek}}{1990}]{PKS0745photometry1}
{Wright} A.,  {Otrupcek} R.,  1990, PKS Catalog (1990, \href
  {https://ui.adsabs.harvard.edu/abs/1990PKS...C......0W} {p.~0}

\bibitem[\protect\citeauthoryear{{You}, {Bickel}  \& {Sokol}}{{You}
  et~al.}{2017}]{You2017}
{You} J.,  {Bickel} C.,   {Sokol} J.,  2017, \mn@doi [Science]
  {10.1126/science.355.6329.1013}, \href
  {https://ui.adsabs.harvard.edu/abs/2017Sci...355.1013Y} {355, 1013}

\bibitem[\protect\citeauthoryear{{Zhuravleva} et~al.,}{{Zhuravleva}
  et~al.}{2014}]{Zhuravleva1410.6485}
{Zhuravleva} I.  et~al., 2014, \mn@doi [\nat] {10.1038/nature13830}, \href
  {https://ui.adsabs.harvard.edu/abs/2014Natur.515...85Z} {515, 85}

\bibitem[\protect\citeauthoryear{{Zhuravleva}, {Churazov}, {Schekochihin},
  {Allen}, {Vikhlinin}  \& {Werner}}{{Zhuravleva}
  et~al.}{2019}]{Zhuravleva1906.06346}
{Zhuravleva} I.,  {Churazov} E.,  {Schekochihin} A.~A.,  {Allen} S.~W.,
  {Vikhlinin} A.,   {Werner} N.,  2019, \mn@doi [Nature Astronomy]
  {10.1038/s41550-019-0794-z}, \href
  {https://ui.adsabs.harvard.edu/abs/2019NatAs...3..832Z} {3, 832}

\bibitem[\protect\citeauthoryear{{van Weeren}, {de Gasperin}, {Akamatsu},
  {Br{\"u}ggen}, {Feretti}, {Kang}, {Stroe}  \& {Zandanel}}{{van Weeren}
  et~al.}{2019}]{vanWeeren2019}
{van Weeren} R.~J.,  {de Gasperin} F.,  {Akamatsu} H.,  {Br{\"u}ggen} M.,
  {Feretti} L.,  {Kang} H.,  {Stroe} A.,   {Zandanel} F.,  2019, \mn@doi [\ssr]
  {10.1007/s11214-019-0584-z}, \href
  {https://ui.adsabs.harvard.edu/abs/2019SSRv..215...16V} {215, 16}

\makeatother
\end{thebibliography}

\appendix

\section{Non-central \lowercase{{\it\bf t}} distribution}
\label{sec:nct}

The non-central $t$ distribution describes a random variable
\begin{equation}
    T = \frac{Z + c}{\sqrt{V/\nu}},
\end{equation}
where $Z$ follows the standard normal distribution and $V$ the $\chi^2$ distribution with $\nu$ degrees of freedom.
The corresponding density function is nontrivial to write in closed form, and is generally evaluated via a relationship with the cumulative distribution function, itself evaluated recursively \citep{Lenth2347693}.
We use the {\sc scipy}\footnote{\url{https://scipy.org/}} implementation of this distribution, which also introduces generic ``location'' and ``scale'' parameters, $\mu$ and $\sigma$, such that $t=(x-\mu)/\sigma$ follows the non-central $t$ distribution.
In practice, we take $x$ to be the difference between $\ln\rxs{}$ and its median 
from fitting a model cluster profile plus 1000 noise realizations.
Thus, the non-central $t$ parameters in Table~\ref{tab:fitresults} describe the sampling distributions of our measurements, i.e.\ the statistical departure of measured values from a model prediction (Equation~\ref{eq:likelihood}).
Note that the limit $c\rightarrow0$ provides a symmetric distribution (the central $t$), while $\nu \gg 1$ approaches the normal distribution (e.g., the distributions for Abell~2029 and Abell~478 are close to normal in $\ln\rxs{}$).

\section{Tabulated proxies for systematics}

Table~\ref{tab:fig3 data} contains the values displayed in Figure~\ref{fig:scaling_factors} and discussed in Section~\ref{sec:systematics}.

\begin{table*}
	\centering
	\caption{Various observable signatures that might be expected to systematically impact the measurement of \rxs{}.
	Column [1] cluster name; 
	[2] projected cluster ellipticity (measured by \citealt{Mantz1502.06020});
	[3] predicted 150\,GHz AGN flux density near 150 GHz extrapolated from lower-frequency measurements (see Section~\ref{sec:radiogalaxies});
	[4] galactic latitude; 
	[5] 100 $\mu$m IRAS surface brightness towards each cluster.
	}
	\label{tab:fig3 data}
	\begin{tabular}{lcccc}
		\hline
		\hline
		Cluster & Ellipticity & AGN flux density & Galactic latitude & IRAS surface brightness \\
		& & (mJy) & (deg) & (MJy sr$^{-1}$) \\
		\hline
        Abell~2029          & $0.198\pm0.014$ & 0.06 &  \phantom{0}$50.5457$ & \phantom{0}3.0 \\
Abell~478           & $0.177\pm0.009$ & 0.45 & $-28.2897$ &           18.1 \\
PKS~0745$-$191      & $0.163\pm0.007$ & 4.47 &   \phantom{00}$3.0299$ &           15.4 \\
Abell~2204          & $0.179\pm0.006$ & 2.65 &  \phantom{0}$33.2374$ & \phantom{0}6.6 \\
RX~J2129.6+0005     & $0.249\pm0.015$ & 0.71 & $-34.4761$ & \phantom{0}3.0 \\
Abell~1835          & $0.122\pm0.007$ & 0.77 &  \phantom{0}$60.5860$ & \phantom{0}2.4 \\
MS~2137.3$-$2353    & $0.128\pm0.013$ & 0.06 & $-46.9371$ & \phantom{0}3.6 \\
MACS~J1931.8$-$2634 & $0.279\pm0.010$ & 8.81 & $-20.0933$ & \phantom{0}5.8 \\
MACS~J1115.8+0129   & $0.244\pm0.015$ & 0.42 &  \phantom{0}$55.6256$ & \phantom{0}3.4 \\
MACS~J1532.8+3021   & $0.196\pm0.017$ & 1.19 &  \phantom{0}$54.6451$ & \phantom{0}2.3 \\
MACS~J1720.2+3536   & $0.185\pm0.017$ & 0.15 &  \phantom{0}$33.0773$ & \phantom{0}2.8 \\
MACS~J0429.6$-$0253 & $0.205\pm0.027$ & 6.70 & $-32.5885$ & \phantom{0}3.8 \\
RX~J1347.5$-$1145   & $0.205\pm0.013$ & 4.39 &  \phantom{0}$48.8076$ & \phantom{0}4.1 \\
MACS~J1423.8+2404   & $0.168\pm0.017$ & 0.76 &  \phantom{0}$68.9856$ & \phantom{0}2.3 \\

		\hline
	\end{tabular}
\end{table*}

\bsp	
\label{lastpage}
\end{document}